\documentclass[gmd, manuscript]{copernicus}

\usepackage{multicol}
\usepackage{multirow}
\usepackage[algo2e,ruled]{algorithm2e}
\usepackage{needspace}
\usepackage{etoolbox}
\makeatletter
\pretocmd{\@startsection}{\Needspace{6\baselineskip}}{}  {\PackageError{manuscript}{Could not install heading page-break protection}{}}
\makeatother

\begin{document}
\nolinenumbers

\title{EBCC v0.2: error-bounded compression for weather and climate applications}

\Author[1]{Langwen}{Huang}
\Author[1,4]{Luigi}{Fusco}
\Author[1]{Florian}{Scheidl}
\Author[2]{Jan}{Zibell}
\Author[2]{Michael Armand}{Sprenger}
\Author[1]{Alexandru}{Calotoiu}
\Author[3]{Sebastian}{Schemm}
\Author[1][torsten.hoefler@inf.ethz.ch]{Torsten}{Hoefler} 
\affil[1]{Scalable Parallel Computing Laboratory, Department of Computer Science, ETH Zurich, Switzerland}
\affil[2]{Institute for Atmospheric and Climate Science, Department of Environmental Systems Science, ETH Zurich, Switzerland}
\affil[3]{Department of Applied Mathematics and Theoretical Physics, Centre for Mathematical Sciences, University of Cambridge, Cambridge, UK}
\affil[4]{Microsoft Ib\'erica, Madrid, Spain}

\runningtitle{EBCC v0.2: error-bounded compression for weather and climate applications}

\runningauthor{L. Huang et al.}

\received{}
\pubdiscuss{} \revised{}
\accepted{}
\published{}

\firstpage{1}

\maketitle

\begin{abstract}
As the resolution of weather and climate simulations increases, the amount of data produced is growing rapidly from hundreds of terabytes to tens of petabytes. These data volumes limit broader adoption, and their continued growth will soon exhaust available storage capacity. To address this problem, we present EBCC v0.2 (Error Bounded Climate-data Compressor). EBCC follows a two-layer compression approach: a base layer using JPEG2000 to capture the bulk of the data at high compression ratios, and a residual layer storing sparse wavelet coefficients to eliminate long-tail extreme errors efficiently. It supports three-dimensional chunks through a multicomponent transform (MCT) in the base layer. To achieve error bounded compression, we introduce error-bound-aware rate allocation in the JPEG2000 base layer. It selects coding passes to approximately satisfy the error target while optimizing the trade-off between compression ratio and mean squared error. We evaluate EBCC alongside established compression methods on benchmarks relevant to weather and climate science, including error statistics, sequential and parallel throughput tests, case studies of primitive and derived variables near a tropical cyclone, an analysis of global energy-budget closure, and a Lagrangian air-parcel trajectory simulation. EBCC either outperforms or closely follows the strongest existing baseline across these benchmarks at relative error targets ranging from 0.01\% to 10\%, achieving compression ratios from 15$\times$ to more than 300$\times$. In the energy-budget-closure and Lagrangian-trajectory benchmarks, EBCC achieves more than 100$\times$ compression while keeping compression-induced errors below an assimilation-uncertainty reference derived from the ERA5 uncertainty ensemble. These results demonstrate that EBCC can produce heavily compressed weather and climate datasets suitable for downstream applications.
\end{abstract}

\introduction  Numerical weather and climate simulations produce a huge amount of data ranging from hundreds of terabytes to tens of petabytes \citep{bauer2021digital, acosta2024computationalcostcmip6, govett2024exascale}. Analyzing these data requires significant storage and computing resources, which are only available to a few researchers.
Moreover, the volume of data is growing rapidly as higher-resolution simulations are required to address the challenges posed by climate change, particularly extreme weather events. As the growth rate of the simulation data eclipses that of storage capacity, a shortage of storage space becomes inevitable \citep{hoefler2023eve, ecmwf2023annual}. As a result, there is an urgent need for data compression to store high-resolution simulation data while ideally having fast data access.

\begin{figure*}[t!]
    \centering
        \includegraphics[width=\textwidth]{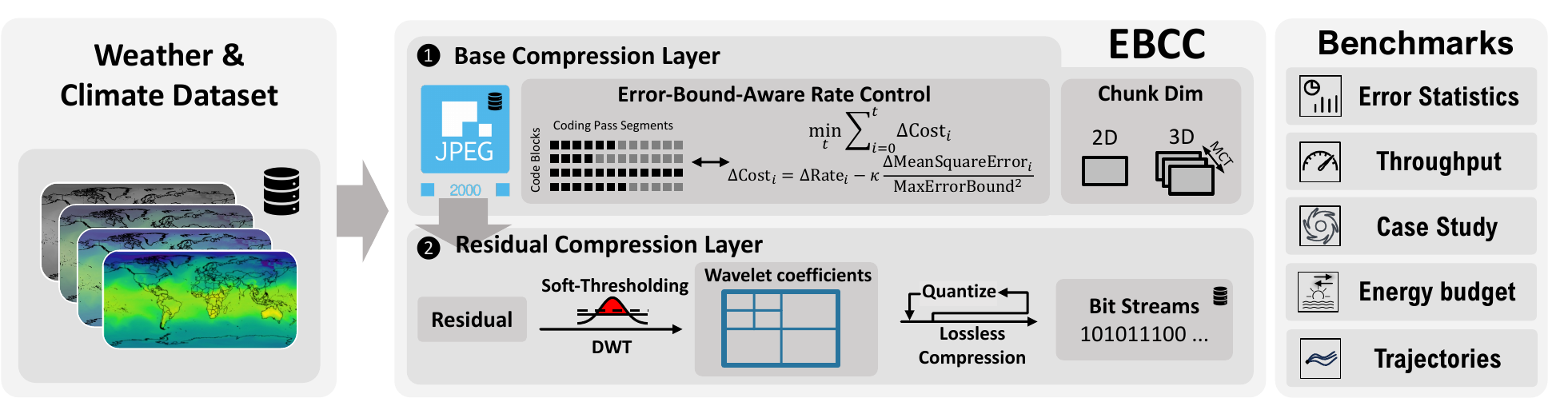}
    \caption{Diagram of the EBCC (Error Bounded Climate Compressor) method. It compresses weather \& climate datasets using a two-layer approach: the base compression layer uses JPEG2000 with error-bound-aware rate allocation for bulk compression, while the residual compression layer packs sparse wavelet coefficients to eliminate long-tail extreme errors efficiently. It supports three-dimensional chunks through a multicomponent transform (MCT) in the base layer. EBCC is evaluated on a suite of weather and climate benchmarks and outperforms or closely follows the strongest existing baseline across a broad range of compression ratios.}
    \label{fig:EBCC_diagram}
\end{figure*}

The exploration of data compression for weather and climate data starts with lossless compression methods \citep{rice1996ccsdslossless, lindstrom2006fpzip, huang2016czip} where data can be exactly restored after compression and decompression processes. However, these methods can hardly reach a high compression ratio above 6$\times$ for common atmospheric variables \citep{mummadisetty2015lossless, huang2016czip} because they attempt to store the random noise component of the data, which is impossible to compress according to Shannon's source coding theorem \citep{shannon1948codetheorem}.

Lossy compression can achieve higher compression ratios than lossless compression by allowing approximate reconstruction of the original data. In scientific applications, compressors that bound the maximum absolute error are commonly used to keep the reconstruction error at every data point within a user-specified tolerance \citep{di2025survey}.

We introduce EBCC (\autoref{fig:EBCC_diagram}), an error-bounded compressor for weather and climate data that prioritizes low RMSE while enforcing a maximum-error bound. It combines error-bound-aware rate allocation in a JPEG2000 base layer with sparse wavelet residual correction to enforce the error bound.

To evaluate EBCC and its effects on downstream applications, we develop a benchmark suite for weather and climate data. It comprises error statistics, sequential and parallel throughput tests, a tropical cyclone case study, an energy-budget closure analysis, and a Lagrangian air-parcel trajectory simulation. The trajectory simulation benchmark, which traces the radioactive cloud released during the Fukushima incident, is highly sensitive to compression errors. We consider it a relevant test for compression methods intended for safety-critical applications.
We evaluate EBCC alongside established error-bounded lossy methods including SZ, SZ3, HPEZ, SPERR, and MGARD. EBCC outperforms or closely follows the strongest existing baseline across a broad range of compression ratios, from 15$\times$ to more than 300$\times$. It concentrates most errors near zero rather than distributing them uniformly within the error bound, thereby reducing distortion in derived variables and simulated trajectories.

\subsection{Related Works}
\label{sec:related_works}

\subsubsection{Lossy compression}

Lossy compression allows for an approximation of the original data in exchange for significantly higher compression ratios. A simple form of lossy compression is uniform scalar quantization, which represents a block of data using linearly scaled integers with fewer bits than the original format. It is often used in conjunction with lossless compression methods to achieve higher compression ratios \citep{silver2017layerpacking}. 
Bit grooming \citep{zender2016bitgrooming} and bit rounding \citep{kouznetsov2021bitrounding} reduce floating-point precision before lossless compression, with bit-information analysis \citep{klower2021bitrounding} guiding the precision selection.
Based on the quantization--lossless compression pipeline, transform-based schemes such as JPEG \citep{jpeg}, JPEG2000 \citep{skodras2001jpeg2000}, wavelet coding with SPECK \citep{pearlman2004speck}, ZFP \citep{lindstrom2014zfp}, and MGARD \citep{gong2023mgard} transform the data before quantization and encoding to improve compression. SPERR \citep{sperr} extends SPECK with an outlier encoding layer to bound the maximum compression error. Learned autoencoder approaches \citep{han2024cra5, mirowski2024healpixcompression, zhao2026weatherlatent} replace predefined transforms with neural encoders and decoders trained on a corpus.
Prediction-based methods estimate values from their neighbors and encode the prediction residuals. FPZIP \citep{lindstrom2006fpzip} uses predictive coding for lossless compression and optionally reduces the retained floating-point precision for lossy compression. SZ \citep{di2016sz,tao2017szquantization} and SZ3 \citep{liang2022sz3} combine prediction with error-controlled quantization to enforce a user-specified error bound. CliZ \citep{jian2024cliz} builds on the SZ3 framework with prediction and encoding tailored to climate datasets. QoZ \citep{liu2022qoz} tunes multilevel interpolation to user-specified quality metrics, while HPEZ (QoZ 2.0) \citep{liu2024hpez} extends this approach with auto-tuned multi-component interpolation.
An alternative approach fits neural representations to individual datasets or images \citep{huang2022nn, ladune2026coolchic}. 

\subsubsection{JPEG2000 related}

Maximum-error control has been studied in JPEG2000 and related wavelet-based compression schemes. \citet{alecu2004deadzone} analyzed embedded deadzone scalar quantizers under maximum-error constraints, and \citet{alecu2006scalable} developed a scalable wavelet coder with an $L_\infty$-oriented embedded bitstream. \citet{lucero2007minimum} searched over lossy JPEG2000 rates and residual quantization precisions to minimize maximum absolute error at a specified compression ratio. A complementary approach combines lossy compression with residual coding: \citet{yea2006twostage} used a wavelet-based lossy layer followed by arithmetic coding of quantized residuals to guarantee a maximum-error bound. For hyperspectral images, \citet{carvajal2008unified} combined a JPEG2000 lossy layer with near-lossless residual coding, while \citet{beerten2015embedded} developed a fully embedded two-stage coder with progressive refinement of both mean-square and maximum absolute error. In climate applications, \citet{woodring2011revisitingjp2k} applied JPEG2000 to ocean-model output and evaluated bit rate against maximum error. EBCC introduces a new approach to bounding JPEG2000 compression errors using error-bound-aware rate allocation and a sparse wavelet residual layer.

\subsubsection{Compression Benchmarks for scientific applications}

There are several studies evaluating rate--error trade-offs in lossy compression of climate data \citep{woodring2011revisitingjp2k,huebbe2013evaluating}. \citet{baker2014evalmethodology,baker2016ensemble,baker2019ssim} developed complementary evaluation approaches spanning statistical comparisons, scientific diagnostics within the CESM Large Ensemble, and SSIM-based visual assessment. \citet{baker2024dssim} extended structural-similarity assessment to floating-point arrays with the data structural similarity index (Data SSIM). \citet{hammerling2019collaborative} and \citet{poppick2020statistical} used spatiotemporal statistical diagnostics to reveal compression artifacts that standard error metrics can miss. These studies motivate evaluation beyond pointwise error bounds or structural-similarity scores alone. SDRBench supplies datasets from multiple scientific domains and common compressor-comparison metrics \citep{zhao2020sdrbench}, while Z-checker provides a framework for statistical, spectral, and error-correlation analyses \citep{tao2019zchecker}. Tools for evaluating compression of weather and climate data include ldcpy for spatiotemporal comparisons \citep{pinard2020ldcpy} and enstools-compression for assessing effects on meteorological analyses \citep{tinto2024enstools}. ClimateBenchPress standardizes comparisons across climate-model and satellite datasets using variable-specific error checks \citep{reichelt2026climatebenchpress}. Application-focused studies also evaluate forecast skill after training on decompressed data \citep{han2024cra5}. Our evaluation of EBCC complements these resources by adding benchmarks inspired by real-world weather and climate research.

\section{Method}
\label{sec:method}

EBCC compresses and decompresses individual chunks of multidimensional scientific data arrays. These chunks are nonoverlapping rectangular subarrays prepared by the calling application or storage layer before they are passed to EBCC. For example, an array indexed by time, vertical level, latitude, and longitude can be partitioned into chunks containing one or more latitude--longitude fields. EBCC processes each supplied chunk independently. We represent each chunk with shape $(C,H,W)$, where $H$ and $W$ are its sizes along the last two input axes (the spatial dimensions), and $C$ is the number of spatial planes. A two-dimensional chunk has $C=1$; for higher-dimensional chunks, only the leading axes are flattened into the component axis $C$, preserving each $H\times W$ plane. EBCC supports two types of error target: maximum absolute error and range-relative maximum absolute error. The former specifies a tolerance $\epsilon>0$ in physical units and requires $\max_i|x_i-\widehat{x}_i|\leq\epsilon$, where $x_i$ and $\widehat{x}_i$ are the original and reconstructed samples in physical units. Following \citet{liang2022sz3}, the range-relative max error is defined as the ratio of the maximum absolute error to the data range in a data chunk (see \autoref{tab:error_metrics} for details).

EBCC uses min--max scaling over all samples in a chunk to map their values to $[0,1]$, then quantizes them to unsigned 16-bit integers (uint16) before feeding them to JPEG2000. The $C$ spatial planes are represented as JPEG2000 components. For $C>1$, we apply a multicomponent transform (MCT) using CDF 9/7 wavelet transform weights. This exploits correlations across, for example, adjacent vertical levels or time steps while preserving the spatial geometry of each plane. We use error-bound-aware rate allocation in this base layer to keep most sample errors within the prescribed bound.

Because the base layer can still leave sparse errors exceeding the bound, we use a residual layer to correct them independently in each $H\times W$ slice. The encoder forms a correction from the base reconstruction errors in physical units and stores it as quantized two-dimensional wavelet coefficients. The decoder reconstructs this correction and adds it to the base reconstruction.

\subsection{Error-bound-aware rate allocation}
JPEG2000 uses embedded block coding with optimized truncation (EBCOT) \citep{taubman2000ebcot} to compress wavelet coefficients and allocate bits across codeblocks to minimize mean squared reconstruction error for a given coded size. EBCOT partitions wavelet subbands into codeblocks and encodes each block independently through successive bit-plane coding passes. Each pass refines the coefficients, producing an embedded bitstream that can be truncated at pass boundaries. The encoder records the cumulative coded size and estimated distortion reduction at each boundary, then selects a prefix of passes from each block. It uses a Lagrange multiplier $\lambda>0$ to trade total coded size $R$ against estimated squared reconstruction error $D$ by minimizing the cost $C=R+\lambda D$ over the  bitstream truncation choices. A longer prefix is preferred when its reduction in distortion, multiplied by $\lambda$, exceeds its additional coded size. Varying $\lambda$ selects operating points on the supported rate--distortion frontier and allows allocation toward a target coded size. EBCC uses this EBCOT allocation mechanism for its base layer, choosing $\lambda$ from the requested pointwise error bound.

For codeblock $k$, let $R_{k,j}$ be the cumulative coded size through pass $j$, and $\Delta D_{k,j}$ the cumulative reduction in squared reconstruction error estimated by the encoder. The empty prefix has $R_{k,0}=\Delta D_{k,0}=0$. If $D_{k,0}$ denotes the initial distortion, the remaining estimate is $D_{k,j}=D_{k,0}-\Delta D_{k,j}$. Defining the block cost as $C_{k,j}=R_{k,j}+\lambda D_{k,j}$, minimizing the total cost $C=\sum_k C_{k,j_k}$ separates into independent block decisions,
\begin{equation}
 j_k^*=\arg\min_j C_{k,j}=\arg\min_j\left[R_{k,j}-\lambda\Delta D_{k,j}\right].
 \label{eq:price_allocation}
\end{equation}
Thus $\lambda$ is a price assigned to squared error: larger prices retain more coding passes.

Let $\epsilon>0$ denote the absolute error tolerance in physical units and $\delta=g\epsilon$ its value on the 16-bit integer scale used by JPEG2000, where $g$ is the chunk-specific scaling factor mapping physical values to this scale. The error bound provides a natural scale for this price. Since $R$ measures coded bitstream size and $D$ has squared-amplitude units, $\lambda$ must have units of coded size per squared amplitude. If amplitudes and their bound are multiplied by a factor $h$, squared errors scale by $h^2$; preserving their relative cost requires $\lambda$ to scale by $h^{-2}$. Based on this dimensional analysis, we select $\lambda$ inversely proportional to the squared error tolerance,
\begin{equation}
 \lambda=\frac{\kappa}{\delta^2},
 \qquad
 C_{k,j}=R_{k,j}+\kappa\frac{D_{k,j}}{\delta^2},
 \label{eq:bound_price}
\end{equation}
where $\kappa$ is a constant factor. A larger $\kappa$ retains more coding passes thus resulting in a lower mean squared error at a fixed error bound $\delta$ and leaving less or eventually no excessive error for the residual layer. Our empirical study in \autoref{sec:kappa_study} finds $\kappa=2.25$ to be a good choice for balancing the base and residual layers to achieve a low overall rate (high overall compression ratio). \autoref{tab:rate_allocation} compares the allocation rules for the three target types. Conventional JPEG2000 allocation searches for a Lagrange multiplier that meets a total rate or distortion target. EBCC instead fixes $\lambda$ directly from the error bound and independently selects the minimum-cost prefix for each codeblock in a single scan. Since EBCC and conventional JPEG2000 use the same coding-pass selection rule, error-bound-aware rate allocation can be viewed as implicitly adapting the base layer's rate and distortion targets to the requested error bound through the choice of $\lambda$.

\begin{table}[htbp]
\small
\centering
\caption{Comparison of JPEG2000 and EBCC base layer rate allocation using the cost $R+\lambda D$. $R_{\mathrm{t}}$ and $D_{\mathrm{t}}$ are the target total coded size and estimated squared-error distortion. $g$ is the chunk-specific scaling factor mapping physical values to the 16-bit integer scale used by JPEG2000, and $\kappa$ is a constant factor. For each codeblock $k$, $j_k^*$ denotes the selected coding-pass prefix.}
\label{tab:rate_allocation}
\renewcommand{\arraystretch}{1.8}
\begin{tabular}{@{}lll@{}}
\tophline
\textbf{Method} & \textbf{Lagrangian multiplier} & \textbf{Coding pass selection}\\
\middlehline
JPEG2000: rate target $R_{\mathrm{t}}$ &
Search $\lambda$ such that $\displaystyle\sum_k R_{k,j_k^*}\leq R_{\mathrm{t}}$ &
\multirow{3}{*}{$\displaystyle j_k^*=\arg\min_j\bigl(R_{k,j}+\lambda D_{k,j}\bigr)$}\\
JPEG2000: distortion target $D_{\mathrm{t}}$ &
Search $\lambda$ such that $\displaystyle\sum_k D_{k,j_k^*}\leq D_{\mathrm{t}}$ &\\
EBCC: max-error target $\epsilon$ &
Set $\lambda=\kappa/(g\epsilon)^2$ directly &\\
\bottomhline
\end{tabular}
\end{table}

\subsection{Sparse wavelet residual correction}
Let $x_i$ denote an original sample, $\widehat{x}^{\mathrm{base}}_i$ its decoded base-layer value in physical units, and $r_i=x_i-\widehat{x}^{\mathrm{base}}_i$ the residual. We encode the required correction using the procedure summarized in \autoref{algo:residual}. If $\max_i|r_i|\leq\epsilon$, no residual payload is needed. Otherwise, we form a soft-thresholded correction with $0<\alpha<1$,
\begin{equation}
 c_i=\operatorname{sign}(r_i)\max\{|r_i|-\alpha\epsilon,0\}.
 \label{eq:residual_deadzone}
\end{equation}
This removes only the excess beyond a deadzone: exact reconstruction of $c$ would give $|r_i-c_i|\leq\alpha\epsilon$. The remaining $(1-\alpha)\epsilon$ is available for residual coding error. We use $\alpha=0.8$ throughout the experiments, selected empirically from small-scale tests to balance correction sparsity and reconstruction quality. Reconstruction quality is examined further in \autoref{sec:ablation_residual}. With this choice, the residual payload remains small across a wide range of the configurations shown in \autoref{fig:method_price}c.

For multicomponent chunks, residual correction and error verification are performed independently on each spatial plane after inverse MCT. We apply a one-level CDF 5/3 wavelet transform to the correction. CDF 5/3 achieved the best overall rate--distortion performance for residual coding in our empirical experiment comparing Haar, CDF 9/7, and CDF 5/3 wavelets. We uniformly quantize the wavelet coefficients with step size $s$. Increasing $s$ coarsens the quantization, producing more zero coefficients and smaller quantized magnitudes, which generally reduces the residual payload size. We then dequantize and inverse-transform the coefficients to obtain the reconstructed correction $\widehat{c}$, which is added to the base reconstruction to obtain the final reconstructed samples $\widehat{x}$. The maximum reconstruction error is bounded by the sum of the deadzone tolerance and the residual coding error,
\begin{equation}
 \|x-\widehat{x}\|_\infty
 \leq\alpha\epsilon+\|c-\widehat{c}\|_\infty.
 \label{eq:residual_bound}
\end{equation}
Our goal is to minimize the residual payload while satisfying the prescribed error bound. To this end, we start with a coarse initial quantization step $s_0$ and reduce it only if reconstruction violates the bound. To choose $s_0$, we use a best-case estimate of the residual coding error relative to the available budget $(1-\alpha)\epsilon$. Rounding introduces at most $s/2$ error per wavelet coefficient, and the inverse transform combines these errors into the reconstructed correction. For the one-level CDF 5/3 transform used here, the best-case ratio of sample-space to coefficient-space maximum error is $1/4$ over the full transform domain, as derived in \autoref{app:cdf53_amplification}. Applying this factor to the coefficient rounding-error scale $s/2$ gives an optimistic sample-space estimate of $s/8$. Matching this estimate to the residual error budget results in $s_0=8(1-\alpha)\epsilon$. We then reconstruct the samples and measure the total error $E(s)=\max_i|x_i-\widehat{x}_i(s)|$ to determine whether $s$ needs refinement.

If $E(s)>\epsilon$, the coefficient step must be refined. To estimate the reduction needed, we use $E(s)-\alpha\epsilon$ as a proxy for the sample-space error introduced by residual coding and assume that it decreases approximately in proportion to $s$. Comparing the available budget $\epsilon-\alpha\epsilon$ with this proxy gives the shrink factor $\rho=(\epsilon-\alpha\epsilon)/[E(s)-\alpha\epsilon]$. We clip this factor to $[0.1,0.8]$ and update the step as $s\leftarrow\rho s$. Since each failed trial reduces $s$ by at least 20\%, coefficient quantization errors tend to zero, so the iteration should eventually satisfy the positive residual error budget. After each update, we repeat quantization and reconstruction with the base stream fixed and explicitly verify the sample-space error. A trial is accepted only if $E(s)\leq\epsilon$; the encoder permits at most $N$ trials and reports failure if none meets this condition.

Once the reconstructed samples satisfy the error bound, the encoder stores the residual as a sequence of nonzero quantized wavelet coefficients. Each coefficient is paired with the number of preceding zeros, allowing the decoder to recover its position without storing the zero coefficients individually. The coefficient values and zero counts are packed as variable-length integers and compressed losslessly using Zstandard (Zstd).

We design the wavelet-based residual compression layer to reduce abrupt changes at the boundaries of regions where the base reconstruction error exceeds the tolerance. In the ablation study in \autoref{sec:ablation_residual}, we compare wavelet-based and direct sample-space residual correction. Wavelet-based correction produces smoother reconstructed fields and spatial gradients while satisfying the same error bound with a similar residual size.

\begin{algorithm2e}[htbp]
\small
\linespread{1.0}\selectfont
\SetAlgoLined
\LinesNumbered
\newcommand{\residualcomment}[1]{\tcp*[r]{\makebox[0.62\linewidth][l]{#1}}}
\KwIn{Original $H\times W$ slice $x$, base reconstruction $\widehat{x}^{\mathrm{base}}$, absolute error bound $\epsilon>0$, deadzone fraction $0<\alpha<1$, and maximum number of trials $N$.}
\KwOut{Residual payload $e_{\mathrm{res}}$ and step $s$, or failure after $N$ unsuccessful trials.}
$r\leftarrow x-\widehat{x}^{\mathrm{base}}$ \residualcomment{Compute the base-layer residual}
\If{$\|r\|_\infty\leq\epsilon$}{\KwRet{empty residual payload}\;}
$c_i\leftarrow\operatorname{sign}(r_i)\max\{|r_i|-\alpha\epsilon,0\}$ \residualcomment{Apply the deadzone, ignore errors below $\alpha\epsilon$}
$w\leftarrow\mathrm{DWT}(c)$ \residualcomment{Apply one-level CDF 5/3 wavelet transform}
$s\leftarrow8(1-\alpha)\epsilon$ \residualcomment{\parbox[t]{0.62\linewidth}{\raggedright Set an optimistic initial step using the best-case\\ sample-space error estimate}}
\For{$t\leftarrow1$ \KwTo $N$}{
 $z\leftarrow\operatorname{round}(w/s)$ \residualcomment{Quantize coefficients}
 $\widehat{c}\leftarrow\mathrm{DWT}^{-1}(sz)$ \residualcomment{Apply inverse wavelet transform}
 $\widehat{x}\leftarrow\widehat{x}^{\mathrm{base}}+\widehat{c}$\;
 $E\leftarrow\|x-\widehat{x}\|_\infty$ \residualcomment{Check the sample-space error}
 \eIf(\hfill\tcp*[f]{\makebox[0.62\linewidth][l]{Error target satisfied}}){$E\leq\epsilon$}{
  $z_p\leftarrow\mathrm{pack}(z)$ \residualcomment{\parbox[t]{0.62\linewidth}{\raggedright Pack nonzero coefficients as variable-length integers\\ and zeros with run-length encoding}}
  $e_{\mathrm{res}}\leftarrow\mathrm{zstd}(z_p)$ \residualcomment{Lossless compression with Zstandard}
  \KwRet{$e_{\mathrm{res}}$, $s$}\;
 }(\hfill\tcp*[f]{\makebox[0.62\linewidth][l]{Error target not satisfied}}){
  $\rho\leftarrow(\epsilon-\alpha\epsilon)/(E-\alpha\epsilon)$ \residualcomment{$0<\rho<1$ since $E>\epsilon$}
  $s\leftarrow s\,\operatorname{clip}(\rho,0.1,0.8)$ \residualcomment{Reduce the step after a failed trial}
 }
}
\KwRet{failure after $N$ unsuccessful trials}\;
\caption{Wavelet residual correction, applied independently to each spatial slice.}
\label{algo:residual}
\end{algorithm2e}

\subsection{Empirical study of $\kappa$ to balance the base and residual layers}
\label{sec:kappa_study}
The scaling in \autoref{eq:bound_price} determines how the squared-error price depends on the error tolerance, but leaves the factor $\kappa$ to be chosen. A small $\kappa$ reduces the base-layer size but can leave more errors for the residual layer to correct, whereas a large $\kappa$ spends more bits in the base layer to reduce that correction cost. Since the base-layer allocation does not account for the resulting residual payload, its rate--distortion objective alone cannot determine which choice minimizes the combined stream size. In addition, the base and residual layers are designed for different working conditions: the base layer is intended to compress the bulk of the input data while leaving sparse excessive errors to be corrected by the residual layer efficiently.

\begin{figure}[t]
\centering
\includegraphics[width=\linewidth]{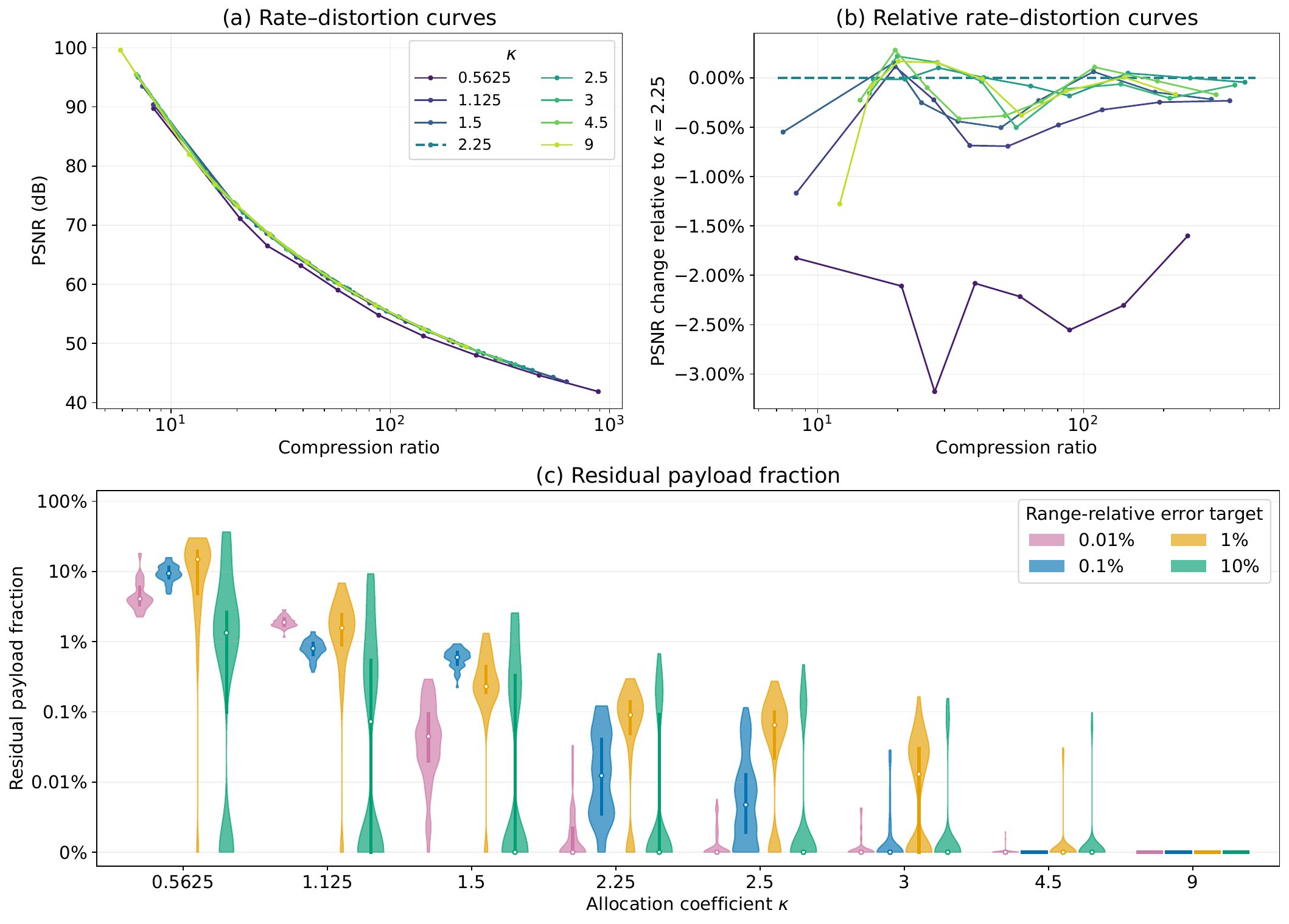}
\caption{Effects of $\kappa$ on EBCC's rate-distortion performance and residual size. Panels show (a) PSNR versus compression ratio (higher is better for both), (b) percentage change in PSNR relative to $\kappa=2.25$ at matched compression ratios, and (c) the fraction of the compressed stream occupied by residual data. Each marker in (a) and (b) corresponds to a range-relative error target from $0.01\%$ to $10\%$. Compression ratios are aggregated using the harmonic mean and PSNR using the arithmetic mean. The reference in (b) is interpolated in log compression ratio without extrapolation; positive values indicate higher PSNR than $\kappa=2.25$. In (c), white dots mark medians, and bars show interquartile ranges.}
\label{fig:method_price}
\end{figure}

In \autoref{fig:method_price}, we sweep $\kappa\in\{0.5625,1.125,1.5,2.25,2.5,3,4.5,9\}$ over ten range-relative error targets from $0.01\%$ to $10\%$ using the same data as the histogram plots in \autoref{sec:basic_stat}. Panels (a) and (b) show that $\kappa=2.25$ achieves the best or close-to-best overall rate--distortion performance across the tested compression ratios. Although some other values yield slightly higher peak signal-to-noise ratio (PSNR) at particular compression ratios, they do not provide a consistent improvement over $\kappa=2.25$.

Panel (c) helps explain this behavior through the allocation of bits between the two layers. Larger values ($\kappa\geq3$) leave little or no residual payload, indicating that the base layer performs nearly all of the compression and underutilizes the residual layer's capacity to correct sparse excessive errors. This additional investment in the base layer does not consistently improve overall rate--distortion performance. Smaller values ($\kappa\leq1.5$) shift more of the correction burden to the residual layer, whose payload occupies a larger fraction of the compressed stream. The resulting overhead generally degrades overall rate--distortion performance, most markedly at $\kappa=0.5625$. Therefore, we use $\kappa=2.25$ in all subsequent experiments to balance the base and residual layers.

\subsection{Implementation details}
We implement EBCC as a C library with a modified OpenJPEG encoder that adds custom rate allocation and a custom multicomponent transform (MCT). EBCC accepts finite single-precision floating-point input and supports absolute and range-relative error bounds. We also provide Python bindings and integrations with HDF5 and Zarr. The HDF5 filter enables transparent decompression in compatible tools including NetCDF, NCO, CDO, and xarray \citep{hdfgroup_hdf5,netcdf4,zender2008nco,schulzweida2023cdo,hoyer2017xarray,zarr}.

In the benchmark, we use the SZ, SZ3, and SPERR compressors shipped by hdf5plugin \citep{hdf5plugin}. We use the default parameters specified in hdf5plugin, except that we set \texttt{swap=True} for SPERR to support C-order HDF5 chunks. We use HPEZ from \url{https://github.com/JLiu-1/HPEZ-QoZ2.0} and enable HPEZ level 4 features with SSIM tuning target. We obtain MGARD from \url{https://github.com/CODARcode/MGARD}. We use LAGRANTO ECMWF 2.7 \citep{sprenger2015lagranto} from \url{https://svn.iac.ethz.ch/websvn/pub/browse/lagranto.ecmwf/tags/2.7/} for Lagrangian trajectory simulations. The versions of the compressors and supporting software are listed in \autoref{tab:compressor_versions} in \autoref{app:software_environment}.

\section{Benchmarks}\label{sec:benchmark}
We design a benchmark suite that ranges from error-statistics analyses on compressed samples to downstream application tests motivated by real-world scenarios. In the first group of benchmarks, we compress a one-year subset of the ERA5 dataset \citep{hersbach2020era5}, measure Baker et al.'s Data SSIM \citep{baker2024dssim} and PSNR (both defined in \autoref{tab:error_metrics}), and examine the rate-distortion relationship of each compression method. Data SSIM scores are averaged arithmetically across time slices, variables, and retained pressure levels. The PSNR values are aggregated using normalized mean squared errors. We also analyze histograms of compression errors and power spectra of decompressed samples at selected compression ratios. Next, we measure sequential compression and decompression throughput and perform a weak-scaling experiment to assess parallel compression throughput. We further examine visual fidelity in a tropical cyclone case study. In addition to directly compressed variables, we compute horizontal divergence from compressed primitive variables to test each compressor's ability to preserve relative values at neighboring locations and across correlated variables. Finally, we apply the compressed data to a global energy-budget closure analysis and a Lagrangian trajectory simulation as representative real-world applications. We also identify error thresholds suitable for practical use by comparing them with the natural variability represented by ERA5 uncertainty members.

We compare EBCC with five baseline error-bounded lossy compressors: SZ, SZ3, HPEZ, SPERR, and MGARD. We vary the range-relative max absolute error tolerance from 0.01\% to 10\% to obtain compressed data with varying compression ratios.
The energy-budget and Lagrangian-trajectory rate-distortion benchmarks use 3D chunks spanning all pressure levels, latitudes, and longitudes at each time; surface variables retain 2D chunks. Each horizontal slice is normalized separately for 3D chunks so that large differences between vertical levels do not obscure smaller variations within a level \citep{silver2017layerpacking}. For 2D chunks, no extra normalization is needed as it can be handled natively by compressors.

\begin{table}[h]
    \small
    \centering
    \caption{Variables used in the benchmarks, sizes of related data stored in 32-bit floating-point format, and compression chunk sizes. Chunk shapes are written as (time, pressure level, latitude, longitude), retaining singleton dimensions; surface variables use a singleton pressure-level dimension. The energy-budget and trajectory rate-distortion benchmarks use all 37 pressure levels for pressure-level variables. \texttt{u850} refers to zonal wind \texttt{u} at pressure level 850hPa, similar for \texttt{v850} and \texttt{u1000}. For energy budget and trajectory simulation benchmarks, only 1 out of 10 ERA5 uncertainty members is compressed.}\label{tab:variable_usage}
    \setlength{\tabcolsep}{4pt}
    \begin{tabular}{lccc}
    \tophline
    \textbf{Benchmark} & \textbf{Variables in Short Name} & \textbf{Chunk Size} & \textbf{Raw Data Size} \\
    \middlehline
    \multirow{2}{*}{Error Statistics} & \multirow{2}{*}{\texttt{z, t, q, u, v, w, u10, v10, t2m, msl}} & \multirow{2}{*}{$(1, 1, 721, 1440)$} & 938.6 MB (Histogram, Spectrum) \\
    & &  & 8.2 TB (Rate-Distortion Plot) \\
    Throughput & \texttt{z, t, q, u, v, w, u10, v10, t2m, msl} & $(1, 1, 721, 1440)$ & 961.1 GB \\
    Case Study & \texttt{u850, v850, u1000, t2m} & $(1, 1, 721, 1440)$ & 15.8 MB \\
    \multirow{2}{*}{Energy Budget} & \texttt{rsnt, rlnt, rlns, rsns, hfss, hfls} & $(1, 1, 361, 720)$ & \multirow{2}{*}{117.8 GB (1 member)} \\
    & \texttt{z, t, q, v} & $(1, 37, 361, 720)$ &  \\
    \multirow{2}{*}{Trajectory Simulation} & \texttt{ps} & $(1, 1, 361, 720)$ & \multirow{2}{*}{5.7 GB (1 member)} \\
    & \texttt{u, v, w} & $(1, 37, 361, 720)$ &  \\
    \bottomhline
    \end{tabular}
\end{table}

\begin{table}[t]
    \centering
    \caption{Abbreviations of variables used in the benchmarks. TOA refers to top of atmosphere.}\label{tab:variable_abbrev}
    \footnotesize
    \setlength{\tabcolsep}{3pt}
    \begin{tabular}{@{}l>{\raggedright\arraybackslash}p{0.25\linewidth}c@{\hspace{1.5em}}l>{\raggedright\arraybackslash}p{0.25\linewidth}c@{}}
    \tophline
    \textbf{Short Name} & \textbf{Long Name} & \textbf{Unit} & \textbf{Short Name} & \textbf{Long Name} & \textbf{Unit} \\
    \middlehline
    \multicolumn{6}{l@{}}{\textbf{Pressure-level variables}} \\
    \texttt{z} & Geopotential & $\text{m}^2\text{s}^{-2}$ & \texttt{u} & Zonal wind & $\text{m}\,\text{s}^{-1}$ \\
    \texttt{t} & Temperature & K & \texttt{v} & Meridional wind & $\text{m}\,\text{s}^{-1}$ \\
    \texttt{q} & Specific humidity & $\text{kg}\,\text{kg}^{-1}$ & \texttt{w} & Vertical velocity & $\text{Pa}\,\text{s}^{-1}$ \\
    \middlehline
    \multicolumn{6}{l@{}}{\textbf{Surface variables}} \\
    \texttt{u10} & 10m zonal wind & $\text{m}\,\text{s}^{-1}$ & \texttt{rlnt} & Net longwave radiation at TOA & $\text{W}\,\text{m}^{-2}$ \\
    \texttt{v10} & 10m meridional wind & $\text{m}\,\text{s}^{-1}$ & \texttt{rlns} & Net longwave radiation at surface & $\text{W}\,\text{m}^{-2}$ \\
    \texttt{t2m} & 2m temperature & K & \texttt{rsns} & Net shortwave radiation at surface & $\text{W}\,\text{m}^{-2}$ \\
    \texttt{msl} & Mean sea level pressure & Pa & \texttt{hfss} & Surface sensible heat flux & $\text{W}\,\text{m}^{-2}$ \\
    \texttt{ps} & Surface pressure & Pa & \texttt{hfls} & Surface latent heat flux & $\text{W}\,\text{m}^{-2}$ \\
    \texttt{rsnt} & Net shortwave radiation at TOA & $\text{W}\,\text{m}^{-2}$ & & & \\
    \bottomhline
    \end{tabular}
\end{table}

\subsection{Error Statistics}\label{sec:basic_stat}

\begin{figure}[htbp]
    \centering
            \includegraphics[width=\linewidth]{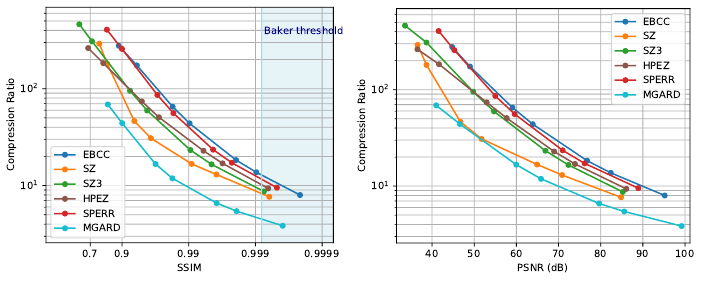}
    \caption{Rate-distortion plot of compression ratio (the higher the better), Data SSIM (left, the higher the better), and PSNR (right, the higher the better) for a subset of the ERA5 dataset (Table \ref{tab:variable_usage}) with varying compression methods and parameters. The 0.99919 SSIM threshold corresponding to the visually indistinguishable regime identified by field experts \citep{baker2024dssim} is marked on the left panel. When calculating Data SSIM, we downsample the original and reconstructed fields to $1^\circ$ resolution so that this threshold is applicable. The numbers are grouped according to the compression method and error target (0.01\%, 0.05\%, 0.1\%, 0.5\%, 1\%, 5\%, 10\%), and aggregated using the harmonic mean for compression ratios and the arithmetic mean for Data SSIM over all variables, retained pressure levels, and timesteps. PSNR values are aggregated using normalized mean squared errors.}
    \label{fig:cr_ssim}
\end{figure}

For the error-statistics benchmark, we compress an 8.2 TB ERA5 subset from 2023 containing 10 pressure-level and surface variables (\autoref{tab:variable_usage}, \autoref{tab:variable_abbrev}). We exclude specific humidity at pressure levels $\leq250$hPa from aggregate quality comparisons. Specific humidity at high altitudes is close to zero, and high-frequency noise present in the original ERA5 specific humidity fields at these altitudes can distort the aggregated SSIM and PSNR metrics at small error bounds for all tested compressors.
Compression ratios use the harmonic mean. The hurricane case study calculates the Data SSIM of data in the plotted region at native resolution.

\autoref{fig:cr_ssim} shows compression ratio versus Data SSIM (left) and PSNR (right) for each compression method at relative error targets ranging from 0.01\% to 10\%, and marks the 0.99919 SSIM threshold corresponding to the visually indistinguishable regime identified by field experts \citep{baker2024dssim}. EBCC generally outperforms the other methods by achieving higher compression ratios at the same Data SSIM or PSNR level, with SPERR providing the closest performance over their common range.

\begin{figure}[h]
    \centering
            \includegraphics[width=\linewidth]{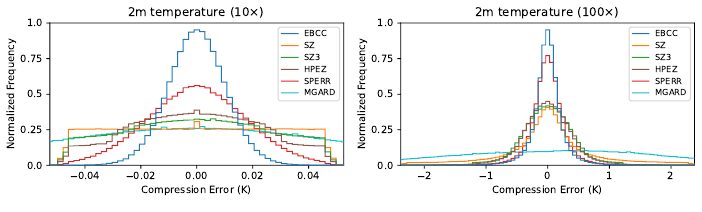}
            \caption{Histogram of compression errors for 2m temperature (\texttt{t2m}) with compression ratios around 10$\times$ (left) and 100$\times$ (right). Each histogram is calculated on $721\times1440$ samples. The x-axis is cropped at the maximum absolute compression error of EBCC.}
    \label{fig:hist_t2m}
\end{figure}

\begin{figure}[h]
\centering
            \includegraphics[width=\linewidth]{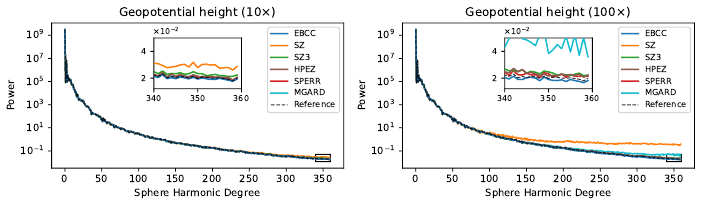}
    \caption{Spherical harmonic power spectra of compressed and original data for geopotential (\texttt{z}) at 500hPa with compression ratios around 10$\times$ (left) and 100$\times$ (right). Each power spectrum is calculated on a $721\times1440$ grid.}
    \label{fig:spectra_z500}
\end{figure}

Furthermore, we examine compression errors in histogram plots (\autoref{fig:hist_t2m}) and power spectra (\autoref{fig:spectra_z500}). In these analyses, we select compressed data with compression ratios close to 10$\times$ and 100$\times$ by searching the error target. For computational efficiency, we restrict this analysis to a single-timestep sample at 00 UTC on 1 March 2023. According to the error histogram of compressed 2m temperature in \autoref{fig:hist_t2m}, EBCC concentrates most errors near zero and produces the narrowest central peak at both target ratios. At a compression ratio of 10$\times$, EBCC yields an approximately Gaussian error distribution closely followed by SPERR, while errors of the other methods are more dispersed. SZ and SZ3 produce approximately uniform error distributions. HPEZ improves upon SZ3 by producing more small errors, but it is worse than EBCC and SPERR. MGARD also produces broader error distributions than EBCC and SPERR. As the compression ratio increases to 100$\times$, EBCC and SPERR retain sharp central peaks, whereas MGARD remains broadly distributed over the displayed range. Histogram plots for other variables are available in the appendix (\autoref{fig:hist_pl_10}--\ref{fig:hist_sfc}) and show a similar trend to \autoref{fig:hist_t2m}.

In the spectra shown in \autoref{fig:spectra_z500}, we compute the power of each spherical-harmonic degree for the original and compressed geopotential at 500hPa. We focus on this variable because the differences among methods are more pronounced, and the same qualitative behavior is observed for other variables (\autoref{fig:spectrum_pl_10}--\ref{fig:spectrum_sfc}). In the high-accuracy regime at a compression ratio of 10$\times$ (left), all methods reproduce the reference spectrum well. In the zoomed high-frequency range, EBCC, HPEZ, and SPERR are closest to the reference, while SZ introduces artificial high-frequency components. In the high-compression regime at 100$\times$ (right), EBCC, HPEZ, SZ3, and SPERR remain close to the reference; by contrast, SZ's artifacts become more pronounced and extend to medium frequencies, and MGARD has excess power at the highest degrees. EBCC slightly underestimates the highest-frequency content, thereby smoothing the data, while HPEZ, SZ3, and SPERR slightly overestimate it.

\subsection{Sequential and Parallel Throughput}\label{sec:throughput}
\begin{figure}[htbp]
    \centering
        \includegraphics[width=\linewidth]{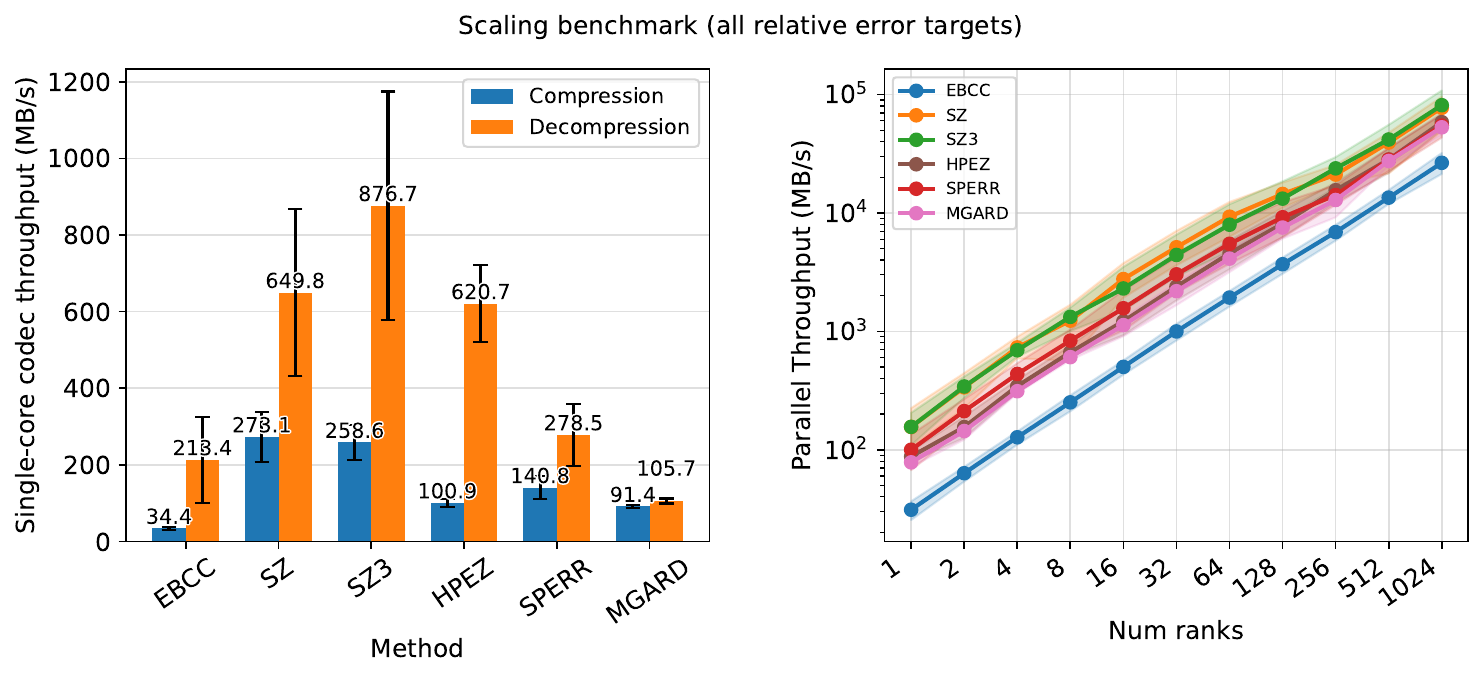}
    \caption{\textbf{Left}: Sequential compression and decompression throughput for each method. Higher is better. Throughput values are aggregated over 226 fields and seven error targets using the harmonic mean. Error bars represent standard deviations across these 1,582 samples. \textbf{Right}: Parallel compression throughput in the weak-scaling experiment. Throughput is computed by dividing the total data size by the elapsed time. The shaded band of each line marks the min--max range across seven error targets, each repeated three times (21 samples in total).}
    \label{fig:throughput}
\end{figure}

We first measure sequential compression and decompression throughput using a single-timestep sample under the same setting as in the previous benchmark. The throughput values are measured on a single thread on NVIDIA GH200 nodes of the Alps supercomputer at the Swiss National Supercomputing Centre (CSCS) \citep{martinasso2025alps}. We preload the data into memory before measuring the throughput to avoid I/O overhead. As shown in \autoref{fig:throughput} (left), EBCC achieves a decompression throughput of 213.4 MB/s, which is comparable to the other methods (105.7--876.7 MB/s). Its compression throughput is lower, at 34.4 MB/s, owing to the overhead of full Tier-1 encoding in the JPEG2000 base layer. However, it can compress multiple data chunks in parallel to achieve higher throughput.

Because weather and climate datasets are typically large and stored on clusters, we perform a more realistic weak-scaling experiment that measures compression throughput for increasing data sizes and increasing numbers of parallel ranks. Unlike the sequential throughput experiment, we measure the end-to-end throughput by dividing total data size by the elapsed time, which accounts for the I/O and other overheads in a practical setting. The experiment runs on the same NVIDIA GH200 nodes, each with 288 CPU cores and 480 GB of memory. We use 1, 2, 4, 8, 16, 32, 64, 128, 256, 512, and 1024 ranks to compress ERA5 subsets with matching numbers of timesteps and sizes ranging from 938.6 MB to 961.1 GB (\autoref{tab:variable_usage}), so that each rank compresses exactly one timestep. We bind each rank to a single CPU core and pack up to 256 ranks onto each node (i.e., jobs with up to 256 ranks run on one node; 512 on 2 nodes; and 1024 on 4 nodes). We limit each node to 256 rather than 288 ranks to avoid CPU contention with CXI network workers under the heavy I/O workload. The results are shown in \autoref{fig:throughput} (right). EBCC scales nearly linearly from 1 rank on 1 node to 1024 ranks on 4 nodes, reaching a parallel compression throughput of 26.5 GB/s. The other methods, including SZ, SZ3, HPEZ, SPERR, and MGARD, reach 52.9--81.3 GB/s at 1024 ranks. The compression-throughput gap between EBCC and SZ3 therefore shrinks from 7.5$\times$ in the sequential setting to 3.1$\times$ in the parallel setting at 1024 ranks on 4 nodes. Based on the weak-scaling results, we estimate that EBCC can compress the entire 2.4 PB ERA5 dataset at 0.25$^{\circ}$ resolution in approximately 25 hours with 4 nodes.

\subsection{Case Study of Hurricane Iota}\label{sec:case_study}
\begin{figure}[h]
    \centering
            \includegraphics[width=\linewidth]{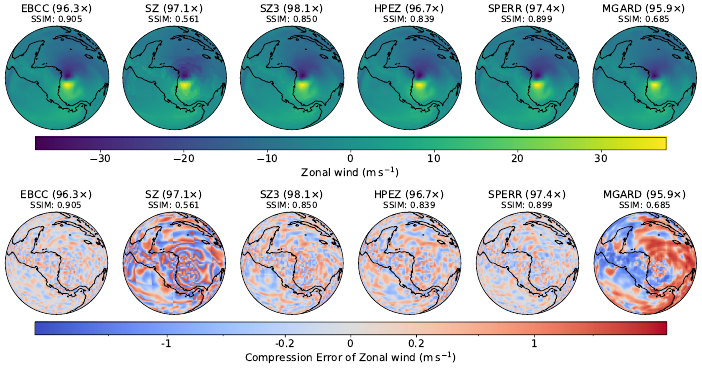}
    \caption{\textbf{Top}: Plots of compressed 850hPa zonal wind speed (u) of Hurricane Iota with compression ratios around 100$\times$. \textbf{Bottom}: Plots of compression errors with corresponding compression ratios as top plots. Achieved compression ratios are marked beside the method names, with the Data SSIM scores calculated on the plotted region.}
    \label{fig:casestudy_u850}
\end{figure}

\begin{figure}[htbp]
    \centering
            \includegraphics[width=\linewidth]{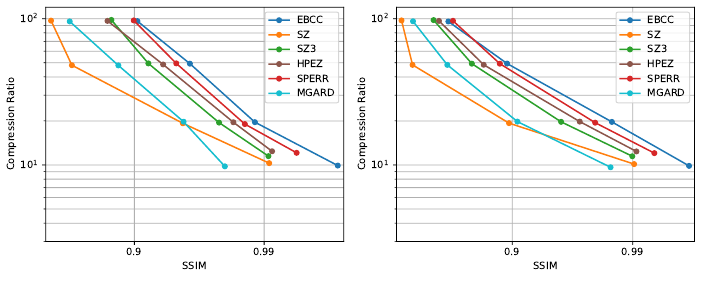}
    \caption{Rate-distortion plots of compression ratio (the higher the better) and Data SSIM (the higher the better) for the Hurricane Iota case study. Left: 850hPa zonal wind. Right: derived 850hPa horizontal wind divergence.}
    \label{fig:casestudy_iota_cr_ssim}
\end{figure}

Because weather and climate data are often inspected visually, compressors must preserve visual fidelity. To test this, we select data containing the record-breaking Hurricane Iota on 16 November 2020. We use binary search to find compression parameters that yield compression ratios of approximately 100$\times$. At 100$\times$ (\autoref{fig:casestudy_u850}), SZ produces obvious block artifacts, MGARD loses fine-scale structure, and SZ3 introduces mild bumps. EBCC and SPERR perform best and do not introduce obvious artifacts. Their leading Data SSIM scores (0.905 and 0.899, respectively) and the residual plots in the bottom row of \autoref{fig:casestudy_u850} support this observation. The rate-distortion plot in \autoref{fig:casestudy_iota_cr_ssim} (left) shows that EBCC achieves higher Data SSIM than the other methods at the same compression ratio over most of the tested range. Additional results for 20$\times$ and 50$\times$ are available in \autoref{fig:casestudy_iota_20}--\ref{fig:casestudy_iota_50}.

\begin{figure}[h]
    \centering
            \includegraphics[width=\linewidth]{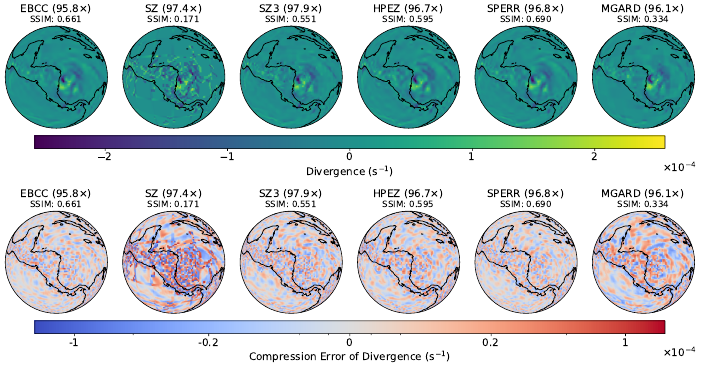}
    \caption{\textbf{Top}: Plots of derived 850hPa horizontal wind divergence from compressed horizontal winds with compression ratios around 100$\times$. \textbf{Bottom}: Plots of compression errors with corresponding compression ratios as top plots. Achieved compression ratios are marked beside the method names, with the Data SSIM scores calculated on the plotted region.}
    \label{fig:casestudy_d850}
\end{figure}

Although EBCC, SZ3, HPEZ, and SPERR preserve the visual features of the directly compressed fields even at 100$\times$, this is less likely for derived variables computed from compressed primitive variables. \autoref{fig:casestudy_d850} shows the derived 850hPa horizontal wind divergence at the hurricane location. Because horizontal divergence depends on the spatial derivatives of the two horizontal wind components, compressors must preserve first-order spatial gradients accurately to maintain fidelity in the computed divergence. At 100$\times$, SPERR achieves the highest Data SSIM score (0.690) closely followed by EBCC (0.661); HPEZ and SZ3 introduce high-frequency spatial artifacts (0.595 and 0.551, respectively); and SZ and MGARD produce strongly degraded results. As the compression ratio decreases, EBCC overtakes SPERR and achieves the highest Data SSIM, as shown in \autoref{fig:casestudy_iota_cr_ssim} (right). The corresponding divergence fields and error plots are shown in the appendix (\autoref{fig:casestudy_iota_20}--\ref{fig:casestudy_iota_50}).

\subsection{Energy Budget Closure Analysis}\label{sec:energy}

\begin{figure}[h]
    \centering
        \includegraphics[width=0.75\linewidth]{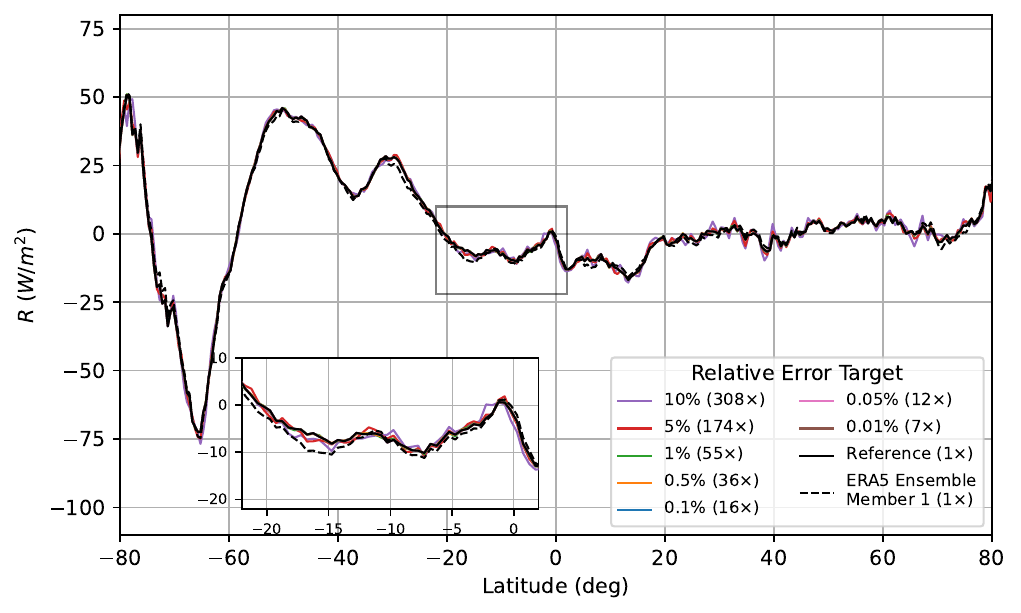}
    \caption{Monthly zonal mean energy budget residual $R$ for compressed and original data for September 2016. The relative error target is set to 0.01\%, 0.05\%, 0.1\%, 0.5\%, 1\%, 5\%, and 10\% respectively. The resulting compression ratios are marked in the legend.}
    \label{fig:energy_budget_delta}
\end{figure}

It is important to maintain the physical consistency of compressed data. In the previous benchmark, we tested a simple derived quantity, horizontal wind divergence. Here, we consider a more demanding benchmark by evaluating the closure of the global atmospheric energy budget using compressed data. In essence, this benchmark measures how compression affects sums and derivatives used to diagnose atmospheric heat transport from three-dimensional atmospheric fields. Ideally, compression errors should be centered around zero so as not to introduce bias into statistics such as spatial or temporal means.
The zonal mean moist static energy (MSE) framework is a commonly used approximation of the atmospheric energy budget which has been used to study the Arctic climate \citep{graversen2016arctic} or storm track shifts \citep{shaw2018moist}. At a given latitude band, there is a balance in the energy fluxes:
\begin{equation}\label{eq:mse_budget}
    R:=\partial_t\langle[\overline{h}]\rangle + \partial_y\langle[\overline{v m}]\rangle - [\overline{F_{NE}}] \approx0
\end{equation}
\citep{Neelin1987ModelingBudget}. The first term on the left-hand side corresponds to the tendency of the thermal energy within the atmospheric column, $h=c_p T+L q$, with $c_p$ the specific heat capacity of dry air and $L$ the latent heat of evaporation, which are both taken as constant here. $T$ and $q$ are air temperature and specific humidity, respectively. The second term denotes the meridional divergence of the flux of MSE that is the product of the meridional wind, $v$, and $m=c_p T+L q + \Phi$ with $\Phi$ the geopotential. The third term represents the net energy (NE) input into the column by surface and top of atmosphere radiative and turbulent heat fluxes. Square brackets denote zonal averages and angle brackets the mass-weighted vertical integration using climatological mean surface pressure \citep{Boer1985VerticallyGlobe}. The overbars denote temporal averaging of the data over a month. Note that the goal is to evaluate the compression with respect to a heat budget diagnostic that is comparatively simple to apply and thus broadly used in the field, for example in \citep{marshall2014the}, instead of more sophisticated approaches \citep{mayer2021consistency}. This comprehensive combination of operations and variables helps us better understand the effect of compression in a real-world scenario.
Using 111 GB of 0.5$^{\circ}$-resolution ERA5 data from August to October 2016, we calculate the terms in \autoref{eq:mse_budget} for September 2016. We then compress the data with relative error targets of 0.01\%, 0.05\%, 0.1\%, 0.5\%, 1\%, 5\%, and 10\%, and compute the zonal-mean energy-budget residual $R$ for the original and compressed data.
Perfect budget closure would correspond to a vanishing residual $R \equiv 0$. The computed residual is approximately $15\,\mathrm{W\,m^{-2}}$ (black line in \autoref{fig:energy_budget_delta}) which is in line with other previous estimates \citep{trenberth2011atmospheric, mayer2017toward} and within the typical range of different turbulent surface heat flux estimates \citep{prince2025constraining}. A considerable fraction of this residual can be linked to uncertainties in assimilated surface heat fluxes \citep{robertson2020uncertainties}. Simplifications in the method following \citep{trenbert1991, mayer2017toward, cox2024new} further introduce discrepancies, particularly over the steep Antarctic topography near $70^{\circ}$\,S. Turning to the evaluation of the compression algorithms, this reference residual serves as a benchmark for the performance of the compression algorithms. In other words, deviations from the reference should ideally be small compared to the reference residual itself.
\begin{figure}[htbp]
    \centering
        \includegraphics[width=0.65\linewidth]{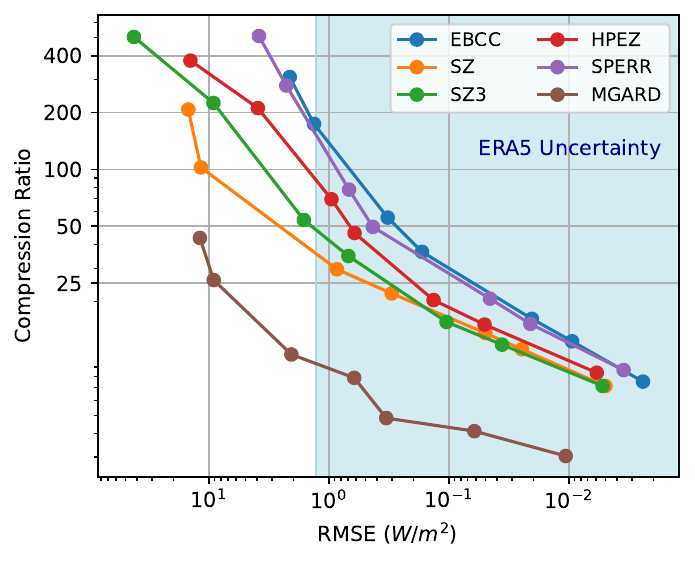}
    \caption{Line plot of compression ratio and RMSE of compressed $R$ (rate-distortion plot). The shaded region indicates RMSE below the smallest RMSE of $R$ from ERA5 uncertainty ensemble members 1--9 relative to member 0.}
    \label{fig:energy_budget_rd_hist}
\end{figure}

As shown in \autoref{fig:energy_budget_delta}, $R$ computed from compressed data remains close to the reference across the tested relative error targets. Even at error targets of 5\% and 10\%, the main zonal structure is preserved at compression ratios of 174$\times$ and 308$\times$, respectively. Quantitatively, we calculate the RMSE of $R$ for the compressed data and plot the resulting rate-distortion relationship for the tested methods (\autoref{fig:energy_budget_rd_hist}). EBCC and SPERR give similar results and generally outperform the other methods in RMSE at comparable compression ratios. The shaded region in \autoref{fig:energy_budget_rd_hist} provides an assimilation-uncertainty reference based on the ERA5 uncertainty ensemble. This provides a practical reference for choosing an error target or compression ratio that keeps compression-induced errors below this uncertainty level. Interpolating the plotted curves in log--log space at this uncertainty reference gives compression ratios of approximately 169$\times$ for EBCC, 153$\times$ for SPERR, 88$\times$ for HPEZ, 48$\times$ for SZ3, 36$\times$ for SZ, and 9.4$\times$ for MGARD. In sum, EBCC and SPERR are the preferable choices to conserve the zonal mean energy budget terms at a monthly scale.

\subsection{Lagrangian Trajectory Simulation}
Finally, we present Lagrangian air-parcel trajectory simulation as a formidable benchmark. Trajectories are highly sensitive to errors in the wind data, and small initial errors amplify over time as the simulator integrates wind speeds at the parcel locations. We design an experiment that simulates air-parcel trajectories around the Fukushima nuclear incident for six days using the Lagrangian analysis tool LAGRANTO \citep{sprenger2015lagranto}, and we examine the trajectory deviations induced by compressed wind data. As in the energy-budget benchmark, we also run the trajectory simulation on ERA5 uncertainty ensemble members to obtain a reference for assimilation uncertainty.

\begin{figure}[htbp]
    \centering
        \includegraphics[width=0.65\linewidth]{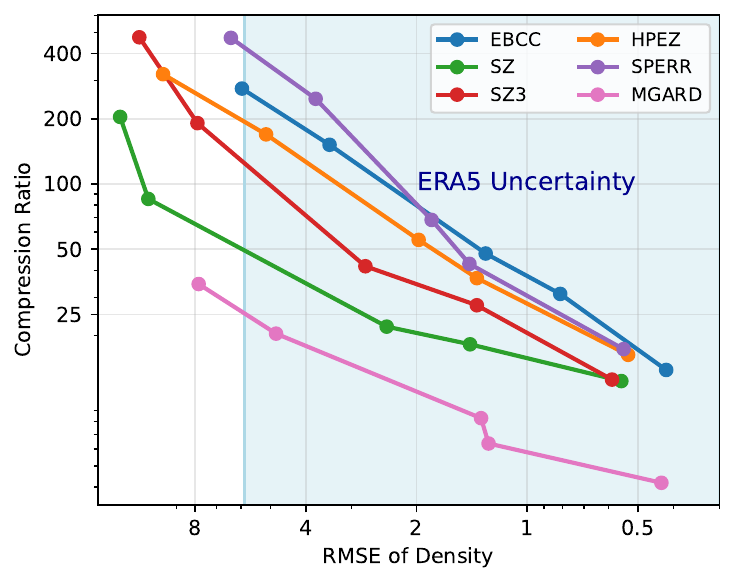}
    \caption{Line plot of compression ratio and RMSE of particle densities at every timestep (rate-distortion plot). Each data point represents one sample of a (compression ratio, RMSE)-pair. The shaded region indicates RMSE below the smallest RMSE of particle densities from ERA5 uncertainty ensemble members 1--9 relative to member 0.}
    \label{fig:trajectory_rmse_rd}
\end{figure}

We measure trajectory deviations using the RMSE of particle densities at every time step, where each tracked air parcel is treated as an abstract particle. This effectively measures the difference between particle distributions.
\autoref{fig:trajectory_rmse_rd} shows the rate-distortion relationship for the tested methods. EBCC and SPERR generally achieve higher compression ratios than the other methods at the same RMSE level and, conversely, lower RMSE at the same compression ratio. EBCC yields the lowest RMSE at lower compression ratios, while SPERR takes the lead at higher compression ratios. The baseline RMSE derived from trajectories computed from the ERA5 uncertainty ensemble is 5.87. This permits a 5\% error target for EBCC, HPEZ, and SPERR, corresponding to compression ratios of 152$\times$, 169$\times$, and 247$\times$, respectively. SZ3 and SZ remain below this threshold at a 1\% target, with compression ratios of approximately 42$\times$ and 22$\times$, while MGARD remains below it at a 5\% target and 20$\times$ compression.

\begin{figure}[htbp]
    \centering
    \includegraphics[width=\linewidth]{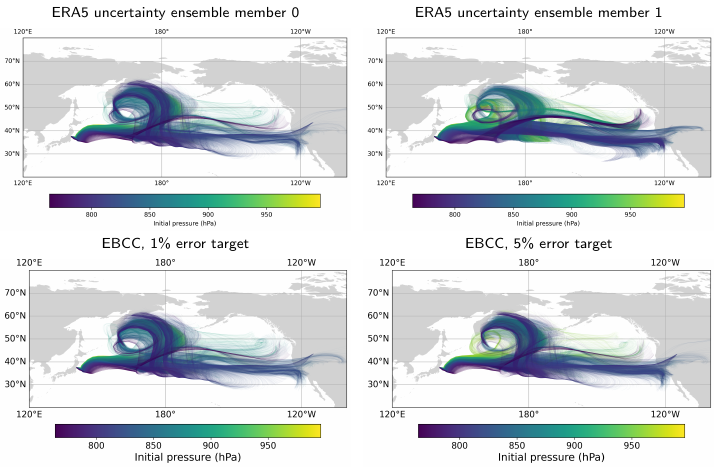}
    \caption{Simulated trajectories of particles around the Fukushima incident over six days, colored by the initial pressure level of the particles. The simulations are performed with original wind data from ERA5 (uncertainty ensemble member 0), ERA5 uncertainty ensemble member 1, and compressed data with EBCC relative error targets of 1\% and 5\% (compression ratios: 48$\times$ and 152$\times$).}
    \label{fig:traj}
\end{figure}

\begin{figure}[htbp]
    \centering
    \includegraphics[width=\linewidth]{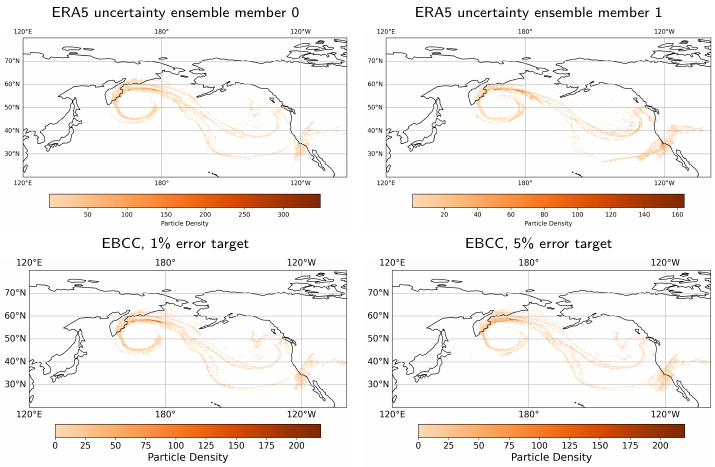}
    \caption{Density of simulated particles one week after the Fukushima incident. The simulations are performed with original wind data from ERA5 (uncertainty ensemble member 0), ERA5 uncertainty ensemble member 1, and compressed data with EBCC relative error targets of 1\% and 5\% (compression ratios: 48$\times$ and 152$\times$). Each simulation contains 35,298 particles.}
    \label{fig:density}
\end{figure}

\autoref{fig:traj} shows simulated trajectories for the original data, an ERA5 uncertainty ensemble member, and compressed data. We select EBCC with relative error targets of 1\% and 5\%, corresponding to compression ratios of 48$\times$ and 152$\times$. Differences between ERA5 uncertainty ensemble members are already evident: in member 1, more lower-level particles circulate over the northwestern Pacific. EBCC with a 1\% relative error target yields a trajectory plume similar to the reference, with only a slight change in the thickness of low-level ($>$ 950hPa) trajectories near 50$^\circ$N, 160$^\circ$E. Relaxing the EBCC error target to 5\% introduces more visible shifts in the trajectories over the northwestern Pacific, while preserving the main transport paths. In \autoref{fig:density}, we examine particle density one week after the Fukushima incident. EBCC at 1\% produces a distribution that is perceptually similar to the reference. At a 5\% error target, EBCC preserves the overall distribution, with local changes in particle concentrations over the North Pacific. 

\section{Ablation Study}
\label{sec:ablation}

We examine how EBCC's design choices affect compression efficiency and reconstruction quality. First, we investigate why EBCC and SPERR differ in performance despite both using wavelet-based compression with residual correction, by varying the base quantization and residual representation. We then examine how wavelet-based and sample-space residual correction affect reconstruction quality with the base reconstruction fixed.

\subsection{Base quantization and residual representation}
EBCC and SPERR achieve similar results in the benchmarks, with EBCC having a slight lead in rate--distortion performance, while both share the approach of wavelet-based compression followed by residual correction. SPERR uses fixed-step quantization of wavelet coefficients followed by sample-space correction of errors exceeding the error target. Default EBCC uses error-bound-aware rate allocation with wavelet-based residual correction. To understand which design choices account for their performance differences, we compare all four combinations of error-bound-aware rate allocation or fixed-step quantization with wavelet-based or sample-space residual correction, alongside SPERR. The fixed-step variants replace coding-pass selection with bound-scaled coefficient quantization and retain all coding passes in the JPEG2000 base layer. We compare the rate--distortion performance of the five configurations using the single-timestep ERA5 sample from the error-statistics benchmark (\autoref{sec:basic_stat}), with range-relative error targets from $0.1\%$ to $10\%$.

\begin{figure}[t]
\centering
\includegraphics[width=\linewidth]{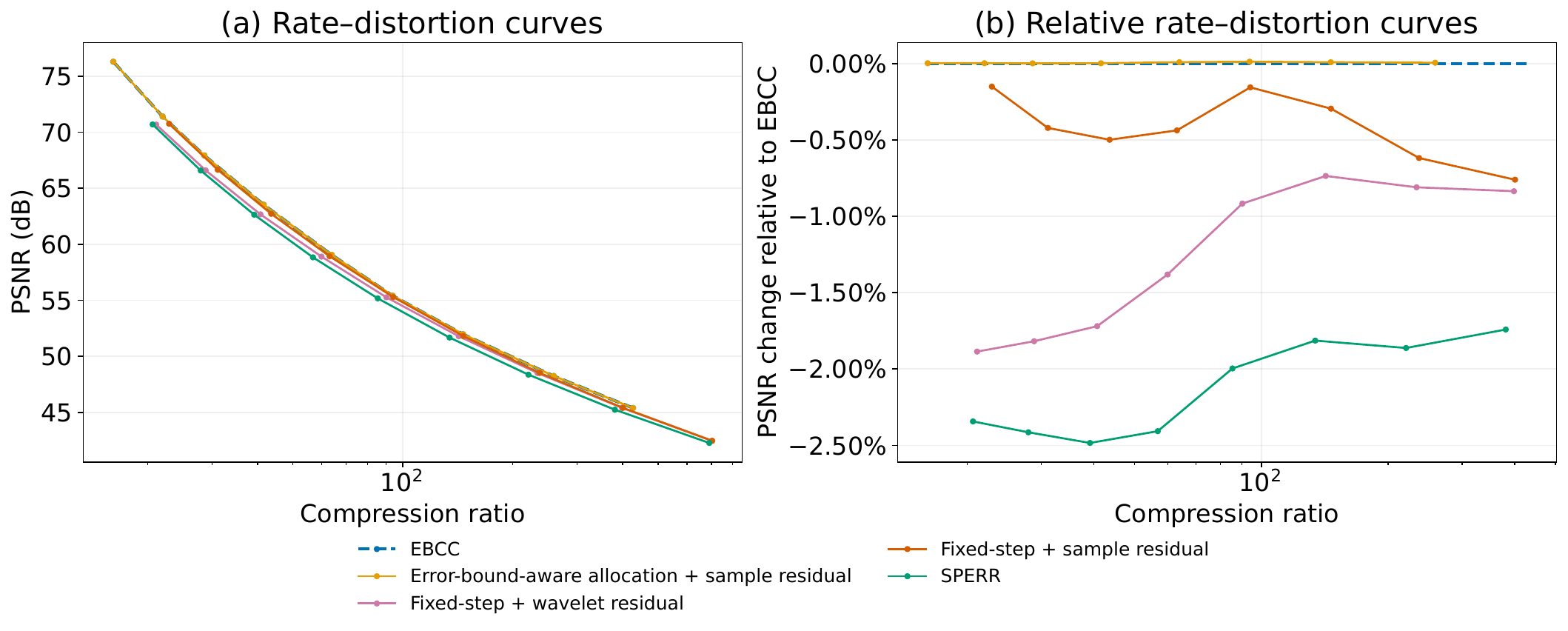}
\caption{Rate-distortion comparison of the EBCC variants and SPERR on the single-timestep ERA5 sample: PSNR versus compression ratio (left, higher is better for both) and percentage change in PSNR relative to EBCC at matched compression ratios (right). Each marker corresponds to a range-relative error target from $0.1\%$ to $10\%$. Compression ratios are aggregated using the harmonic mean and PSNR using the arithmetic mean. The EBCC reference is interpolated in log compression ratio without extrapolation. The dashed line in the right panel marks zero change; positive values indicate higher PSNR than EBCC.}
\label{fig:ablation_quantization}
\end{figure}

In \autoref{fig:ablation_quantization}, error-bound-aware rate allocation achieves higher aggregate PSNR than fixed-step quantization at matched compression ratios with either residual representation. With error-bound-aware allocation, the sample-space and wavelet-based residual variants have nearly identical rate--distortion curves. Under fixed-step quantization, sample-space correction achieves higher aggregate PSNR than wavelet-based correction over the plotted comparison range. These results identify base-layer rate allocation/quantization as the main contributor among the two tested design choices to EBCC's advantage. The EBCC variant using fixed-step quantization with sample-space residual correction also outperforms SPERR. Differences between SPERR's SPECK coding and EBCC's EBCOT-based entropy coding likely contribute to this remaining gap.

\subsection{Effects of residual correction on reconstruction quality}
\label{sec:ablation_residual}

\begin{figure}[p]
\centering
\includegraphics[width=\linewidth]{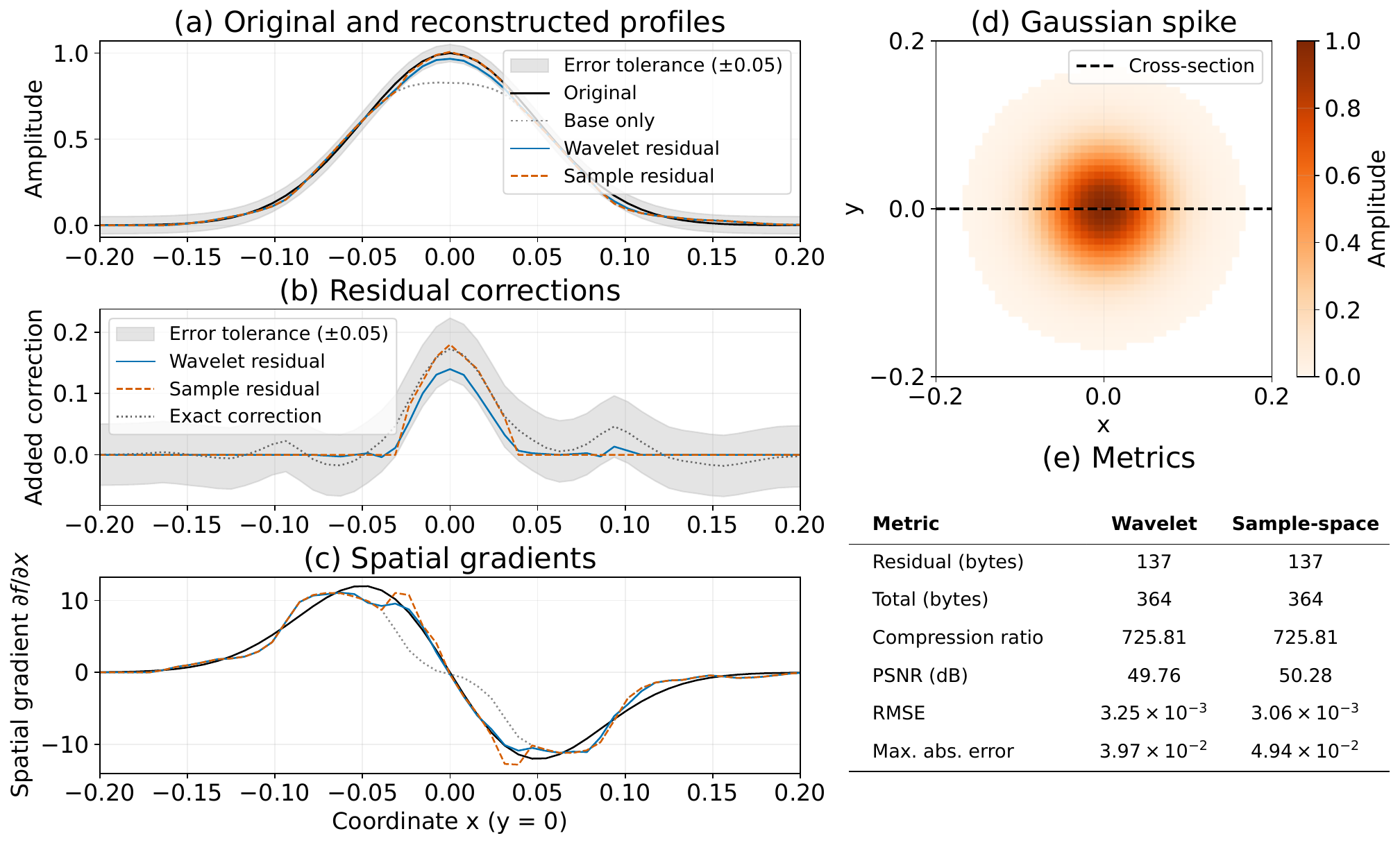}
\caption{Comparison of wavelet-based and sample-space residual correction for an image with a localized Gaussian spike. Panels show (a) original and reconstructed profiles, (b) residual corrections, (c) spatial gradients, (d) original image, and (e) compression and reconstruction metrics. The dashed line in (d) marks the cross-section used in (a)--(c). Shaded bands in (a) and (b) indicate the error tolerance $\pm0.05$ around the original profile and exact residual correction, respectively.}
\label{fig:method_spike}
\end{figure}

The preceding ablation shows a PSNR loss with wavelet-based residual correction compared with sample-space residual correction. To examine their effects on reconstruction quality in more detail, we compare the two forms of residual correction on an image with a localized ``spike'' (\autoref{fig:method_spike}d), defined by $f(x,y)=\exp[-(x^2+y^2)/(2\sigma^2)]$ with $\sigma=0.05$, sampled on a $257\times257$ uniform grid over $(x,y)\in[-1,1]^2$. We reduce $\kappa$ from 2.25 to approximately 0.0275 at an error target of $\epsilon=0.05$ to increase the base reconstruction error and make the residual layer's contribution more pronounced. We match the total compressed size for the two methods.

In \autoref{fig:method_spike}, we examine the reconstructed profiles and spatial gradients along a cross-section where the residual correction is substantial. Both methods restore the peak suppressed by the base reconstruction (\autoref{fig:method_spike}a) and satisfy the same maximum-error tolerance at equal storage cost. Sample-space residual correction follows the peak more closely, whereas wavelet-based residual correction leaves a small underestimate of its amplitude. This is consistent with the slightly higher PSNR of sample-space correction (50.28 versus 49.76 dB). However, its correction changes more abruptly (\autoref{fig:method_spike}b), producing visible kinks and local overshoots in the spatial gradient (\autoref{fig:method_spike}c). Wavelet-based correction distributes the adjustment across neighboring samples, giving smoother gradient transitions, although neither method exactly reproduces the reference gradient. This example shows that wavelet-based residual correction can produce smoother spatial gradients despite a slightly lower PSNR.

\conclusions  This study introduces EBCC, an error-bounded compression method for weather and climate data. EBCC consists of a base compression layer using JPEG2000 to capture the bulk of the data and a residual layer storing sparse wavelet coefficients to correct errors exceeding the given error bound. It supports 3D chunks through a multicomponent transform (MCT) in the base layer. We also introduce error-bound-aware rate allocation in the JPEG2000 base layer. It selects coding passes to approximately satisfy the error target while optimizing the trade-off between compression ratio and mean squared error, with residual correction enforcing the given error bound. EBCC emphasizes minimizing RMSE while providing a maximum absolute error guarantee. We implement EBCC as a standalone C library with Python bindings and HDF5 and Zarr integrations for use in existing toolchains.

We evaluate EBCC on a benchmark suite comprising error statistics, sequential and parallel throughput tests, case studies of compressed primitive and derived variables near a tropical cyclone, a global energy-budget-closure analysis, and a Lagrangian air-parcel trajectory simulation. EBCC either outperforms or closely follows the strongest existing baseline across these benchmarks at relative error targets ranging from 0.01\% to 10\%, achieving compression ratios from 15$\times$ to more than 300$\times$. EBCC scales nearly linearly in an embarrassingly parallel chunk-wise compression setting on GH200 nodes. Based on the weak scaling experiment, compressing the full 2.4 PB ERA5 dataset at 0.25$^{\circ}$ resolution would require approximately 25 hours on 4 nodes. EBCC can reconstruct derivatives of the compressed data with high fidelity. At high compression ratios, EBCC closely follows the spectrum of the reference data, slightly attenuating high-frequency power without introducing spurious power. In the energy-budget and Lagrangian-trajectory benchmarks, EBCC achieves compression ratios above 100$\times$ while keeping compression-induced errors below an assimilation-uncertainty reference derived from the ERA5 uncertainty ensemble.
EBCC is an important step forward to cope with the growing difficulties in storing weather and climate data.

\codedataavailability{EBCC v0.2 is available under the MIT licence at \url{https://github.com/spcl/EBCC}. The source code and benchmark suite, including configurations, data-retrieval scripts, and instructions and scripts to reproduce the figures, are archived on Zenodo \citep{huang_2026_benchmark}.}

\clearpage
\appendix

\section{Error metrics}\label{app:error_metrics}
\autoref{tab:error_metrics} summarizes the error metrics used throughout this study. $A$ and $A'$ denote original and reconstructed arrays with $n$ elements. For range-relative maximum absolute error, the arrays and their extrema refer to a single compression chunk. PSNR uses the original data range and is expressed in decibels.

For Data SSIM, each pair of horizontal fields is jointly scaled to $[0,1]$ and quantized to 256 levels to obtain $X$ and $Y$. The local means $\mu$, variances $\sigma^2$, and covariance $\sigma_{XY}$ use an $11\times11$ Gaussian window with standard deviation 1.5 grid cells; $C_1=C_2=10^{-8}$. The set $\Omega$ excludes a five-cell border. Scores are averaged across horizontal fields. Higher PSNR and Data SSIM indicate better fidelity.

\suppressfloats[t]
\begin{table}[htbp]
\centering
\small
\caption{Error metrics and their mathematical definitions.}\label{tab:error_metrics}
\renewcommand{\arraystretch}{1.8}
\begin{tabular}{@{}p{0.36\linewidth}p{0.59\linewidth}@{}}
\tophline
\textbf{Metric} & \textbf{Formula} \\ \middlehline
Maximum absolute error &
$\displaystyle\max_{1\leq i\leq n}|A'_i-A_i|$ \\
Range-relative maximum absolute error &
$\displaystyle\frac{\max_{1\leq i\leq n}|A'_i-A_i|}{\max_i A_i-\min_i A_i}$ \\
Root mean square error (RMSE) &
$\displaystyle\sqrt{\frac{1}{n}\sum_{i=1}^n(A'_i-A_i)^2}$ \\
Peak signal-to-noise ratio (PSNR) &
$\displaystyle20\log_{10}\!\left(\frac{\max_i A_i-\min_i A_i}{\mathrm{RMSE}}\right)$ \\
Data structural similarity index (Data SSIM) \citep{baker2024dssim} &
$\displaystyle\frac{1}{|\Omega|}\sum_{i\in\Omega}
\frac{(2\mu_{X,i}\mu_{Y,i}+C_1)(2\sigma_{XY,i}+C_2)}
{(\mu_{X,i}^2+\mu_{Y,i}^2+C_1)(\sigma_{X,i}^2+\sigma_{Y,i}^2+C_2)}$ \\
\bottomhline
\end{tabular}
\end{table}

\clearpage
\section{Best-case inverse CDF 5/3 wavelet transform amplification}
\label{app:cdf53_amplification}
We derive the best-case maximum-error ratio for the one-level CDF 5/3 transform used in the residual layer. Let $T$ be the linear forward transform on the full spatial domain, $q$ the coefficient quantization error, and $e=T^{-1}q$ the resulting sample-space error. The best-case inverse-transform amplification factor is
\begin{equation}
 \gamma_{\min}=\min_{q\ne0}\frac{\|T^{-1}q\|_\infty}{\|q\|_\infty}
 =\left(\max_{e\ne0}\frac{\|Te\|_\infty}{\|e\|_\infty}\right)^{-1}
 =\frac{1}{\|T\|_\infty},
 \label{eq:app_min_amplification}
\end{equation}
where the induced matrix norm $\|T\|_\infty$ is the largest absolute row sum. The equality follows by substituting $q=Te$, since $T$ is invertible.

The one-dimensional CDF 5/3 filters are \citep{adams2000reversible,skodras2001jpeg2000}
\begin{equation}
 h_{\mathrm{L}}=\left(-\frac18,\frac14,\frac34,\frac14,-\frac18\right),
 \qquad
 h_{\mathrm{H}}=\left(-\frac12,1,-\frac12\right).
 \label{eq:app_cdf53_filters}
\end{equation}
Index $h_{\mathrm{L}}$ from $-2$ to $2$ and $h_{\mathrm{H}}$ from $-1$ to $1$, with zero values outside these supports. For a one-dimensional array of $M$ input samples, let $T_M$ be the $M\times M$ transform matrix, with low-pass coefficients followed by high-pass coefficients. Filtering and downsampling by two give the interior matrix entries
\begin{equation}
 \begin{aligned}
 (T_M)_{r,n}&=h_{\mathrm{L}}[n-2r],
 &&0\leq r<\lceil M/2\rceil,\quad 0\leq n<M,\\
 (T_M)_{r,n}&=h_{\mathrm{H}}[n-2(r-\lceil M/2\rceil)-1],
 &&\lceil M/2\rceil\leq r<M,\quad 0\leq n<M,
 \end{aligned}
 \label{eq:app_cdf53_matrix}
\end{equation}
Thus each row is a shifted copy of one filter. For six input samples with periodic boundaries, the transform matrix is
\begin{equation}
 T_6=
 \left[\begin{array}{rrrrrr}
  3/4 & 1/4 & -1/8 & 0 & -1/8 & 1/4\\
  -1/8 & 1/4 & 3/4 & 1/4 & -1/8 & 0\\
  -1/8 & 0 & -1/8 & 1/4 & 3/4 & 1/4\\ \hline
  -1/2 & 1 & -1/2 & 0 & 0 & 0\\
  0 & 0 & -1/2 & 1 & -1/2 & 0\\
  -1/2 & 0 & 0 & 0 & -1/2 & 1
 \end{array}\right],
 \label{eq:app_cdf53_matrix_example}
\end{equation}
where the upper three rows produce the low-pass coefficients and the lower three produce the high-pass coefficients. The absolute row sums are $3/2$ for the low-pass rows and $2$ for the high-pass rows, so $\|T_M\|_\infty=2$. The symmetric boundaries used in EBCC preserve this maximum row sum.

The two-dimensional transform applies these filters along rows and columns. Each resulting filter coefficient is a product of the corresponding horizontal and vertical coefficients, so the absolute filter sums multiply. The high--high filter has the largest sum, giving
\begin{equation}
 \|T\|_\infty=2\times2=4,
 \qquad \gamma_{\min}=\frac14.
 \label{eq:app_cdf53_gain}
\end{equation}
Therefore, the maximum absolute sample-space error $\|e\|_\infty$ is at least $1/4$ of the maximum absolute coefficient quantization error $\|q\|_\infty$.

\clearpage
\section{Extended benchmark results}\label{app:si}
\subsection{Histograms}\label{app:histograms}
Histograms use a single-timestep sample at 00 UTC on 1 March 2023. The plotted error range is limited to the maximum absolute EBCC error for each field.
\begin{figure}[htbp]
\centering
\includegraphics[width=\linewidth]{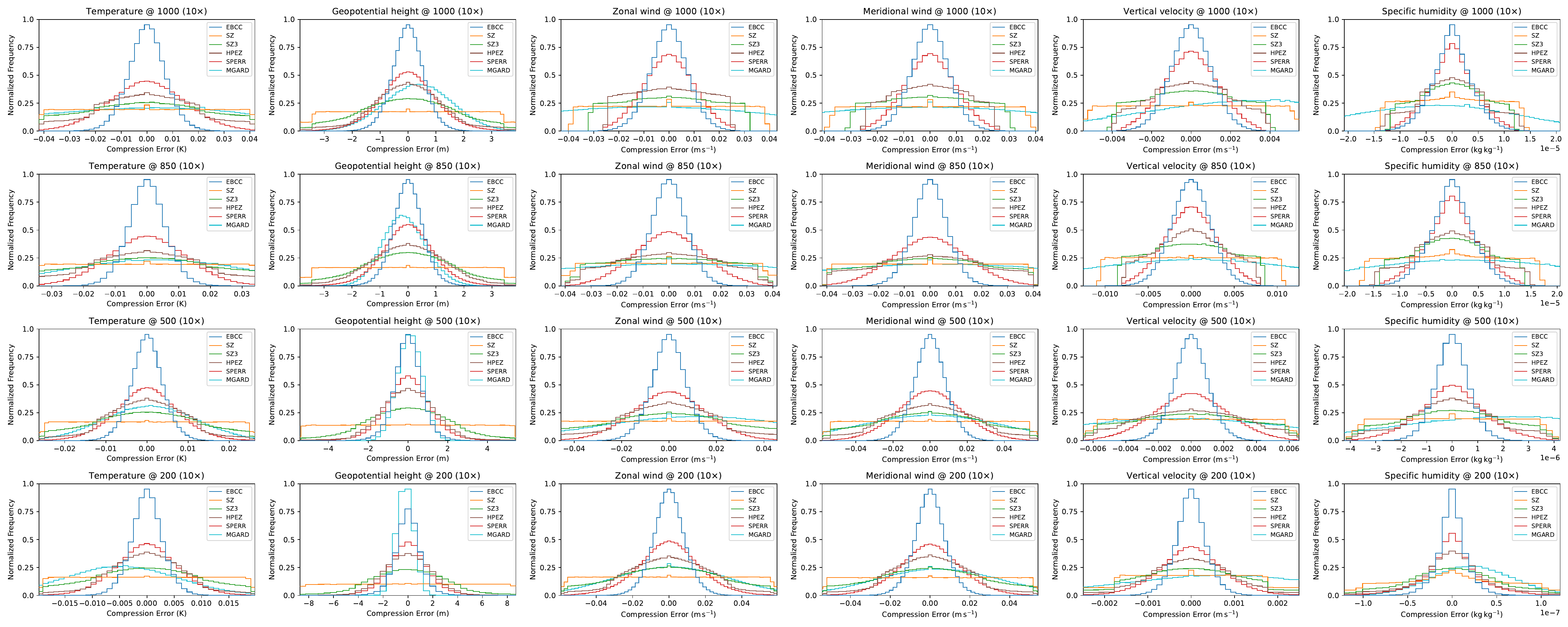}
\caption{Histograms of compression errors for pressure-level variables at a compression ratio of 10$\times$.}\label{fig:hist_pl_10}
\end{figure}

\begin{figure}[htbp]
\centering
\includegraphics[width=\linewidth]{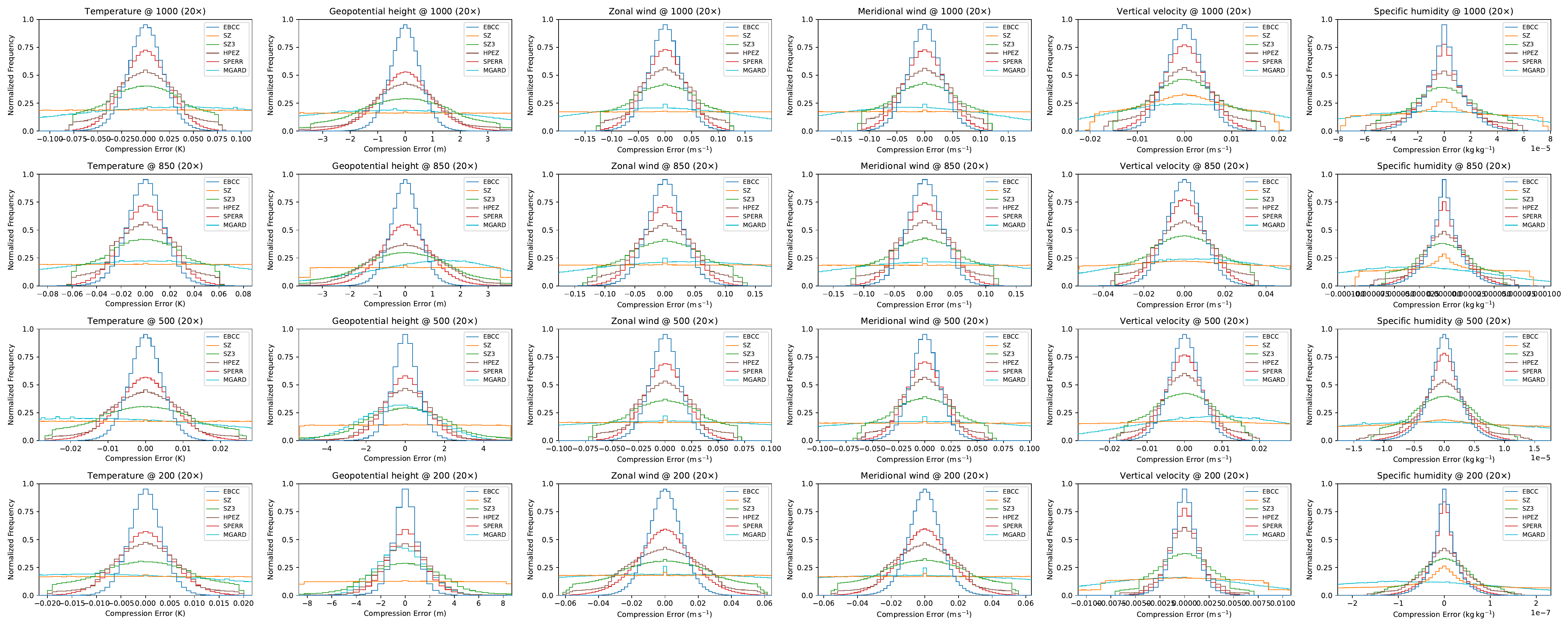}
\caption{Histograms of compression errors for pressure-level variables at a compression ratio of 20$\times$.}\label{fig:hist_pl_20}
\end{figure}

\begin{figure}[htbp]
\centering
\includegraphics[width=\linewidth]{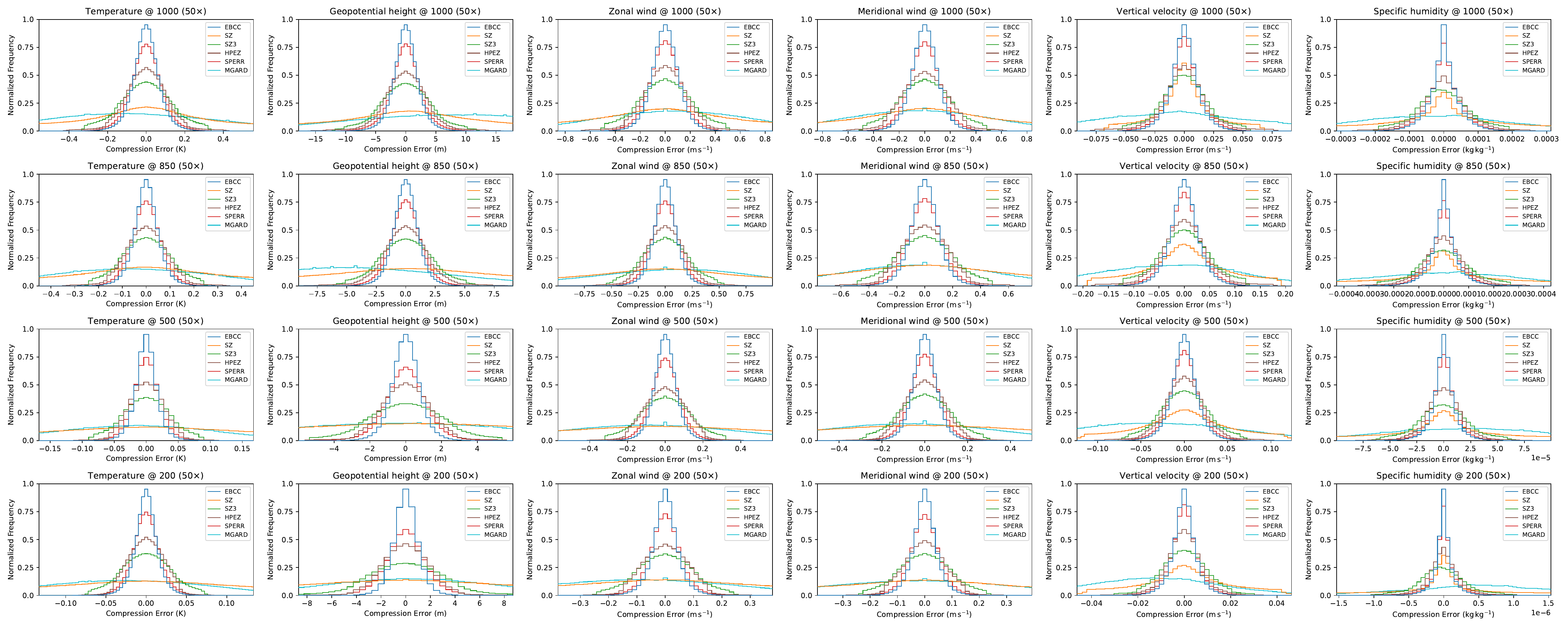}
\caption{Histograms of compression errors for pressure-level variables at a compression ratio of 50$\times$.}\label{fig:hist_pl_50}
\end{figure}

\begin{figure}[htbp]
\centering
\includegraphics[width=\linewidth]{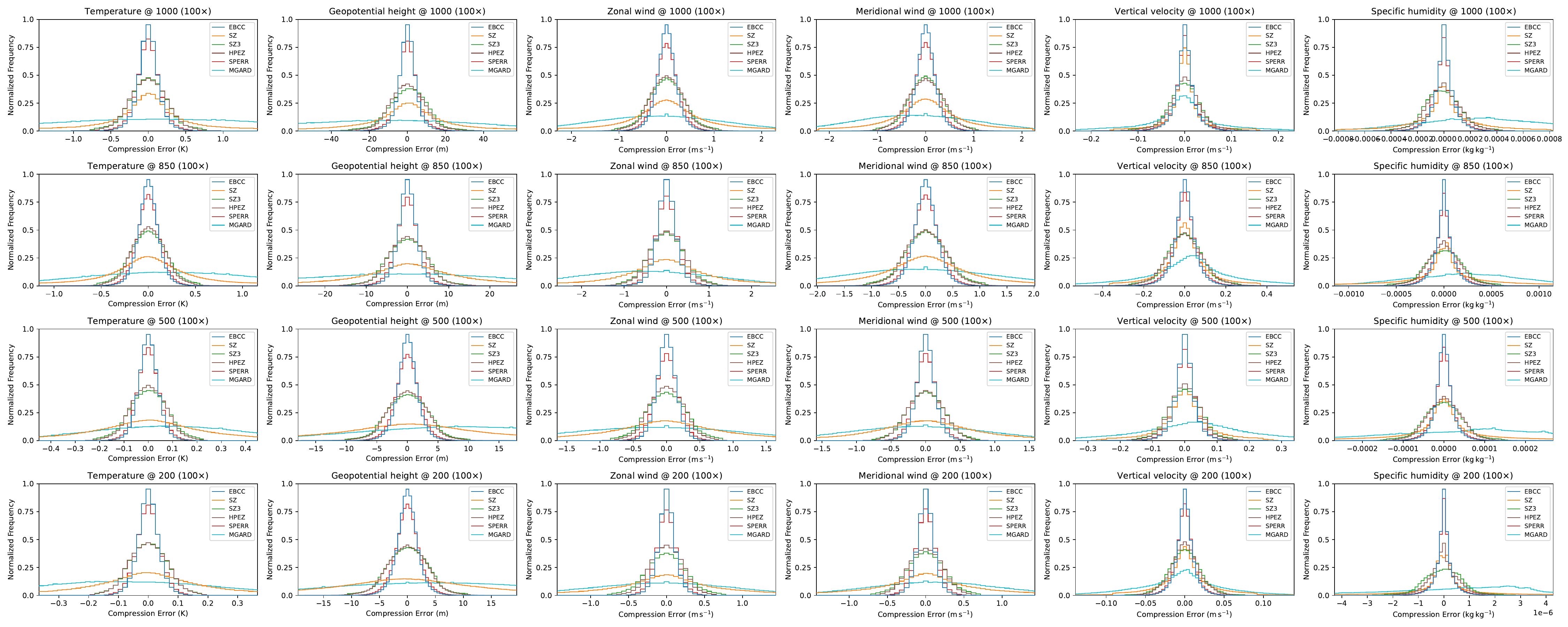}
\caption{Histograms of compression errors for pressure-level variables at a compression ratio of 100$\times$.}\label{fig:hist_pl_100}
\end{figure}

\begin{figure}[htbp]
\centering
\includegraphics[width=\linewidth]{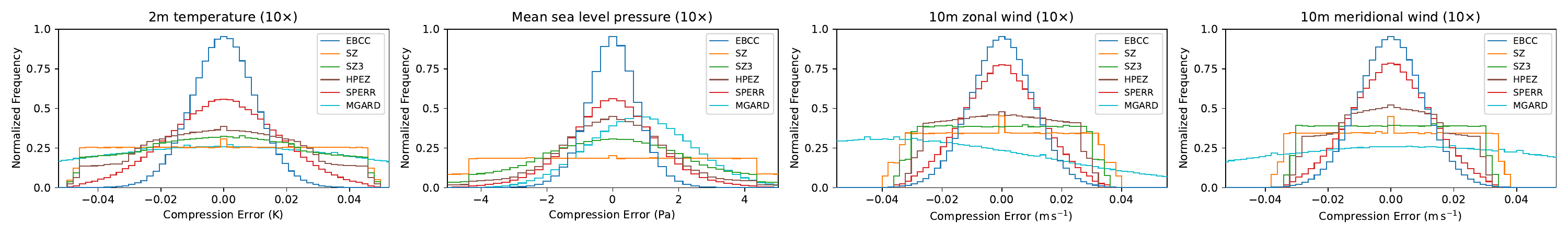}
\includegraphics[width=\linewidth]{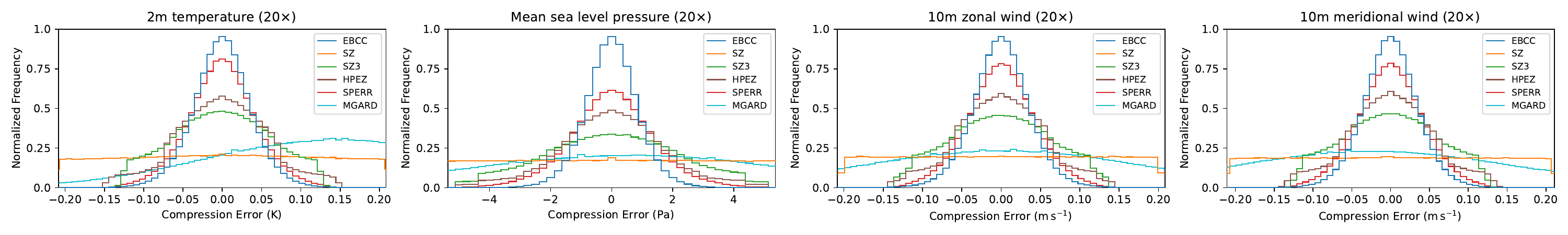}
\includegraphics[width=\linewidth]{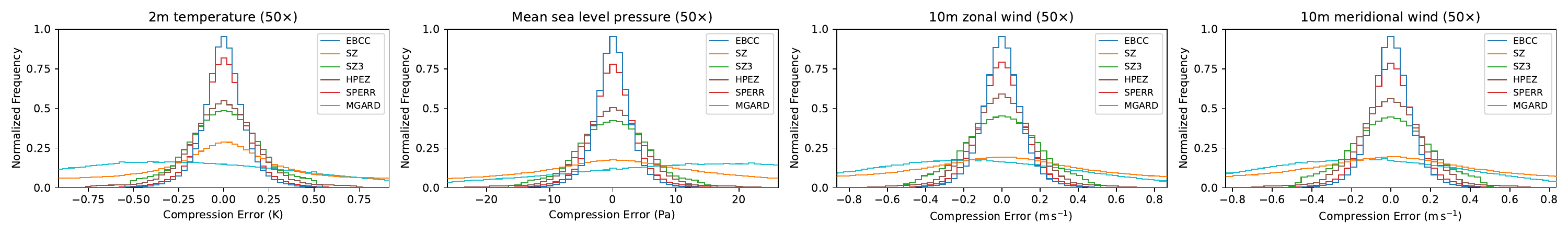}
\includegraphics[width=\linewidth]{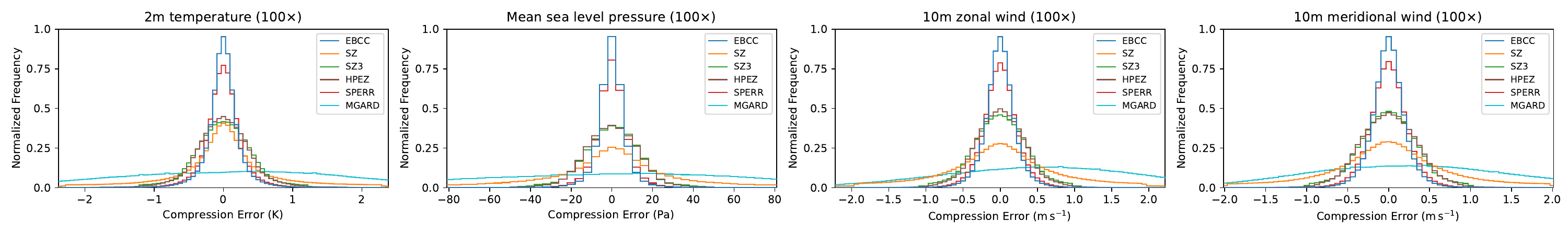}
\caption{Histograms of compression errors for surface variables at compression ratios of 10$\times$, 20$\times$, 50$\times$, and 100$\times$ from top to bottom.}\label{fig:hist_sfc}
\end{figure}

\clearpage
\subsection{Power Spectra}\label{app:spectra}
Spectra and compression ratios are evaluated using the same single-timestep sample at 00 UTC on 1 March 2023 as the histograms in \autoref{app:histograms}.
\begin{figure}[htbp]
\centering
\includegraphics[width=\linewidth]{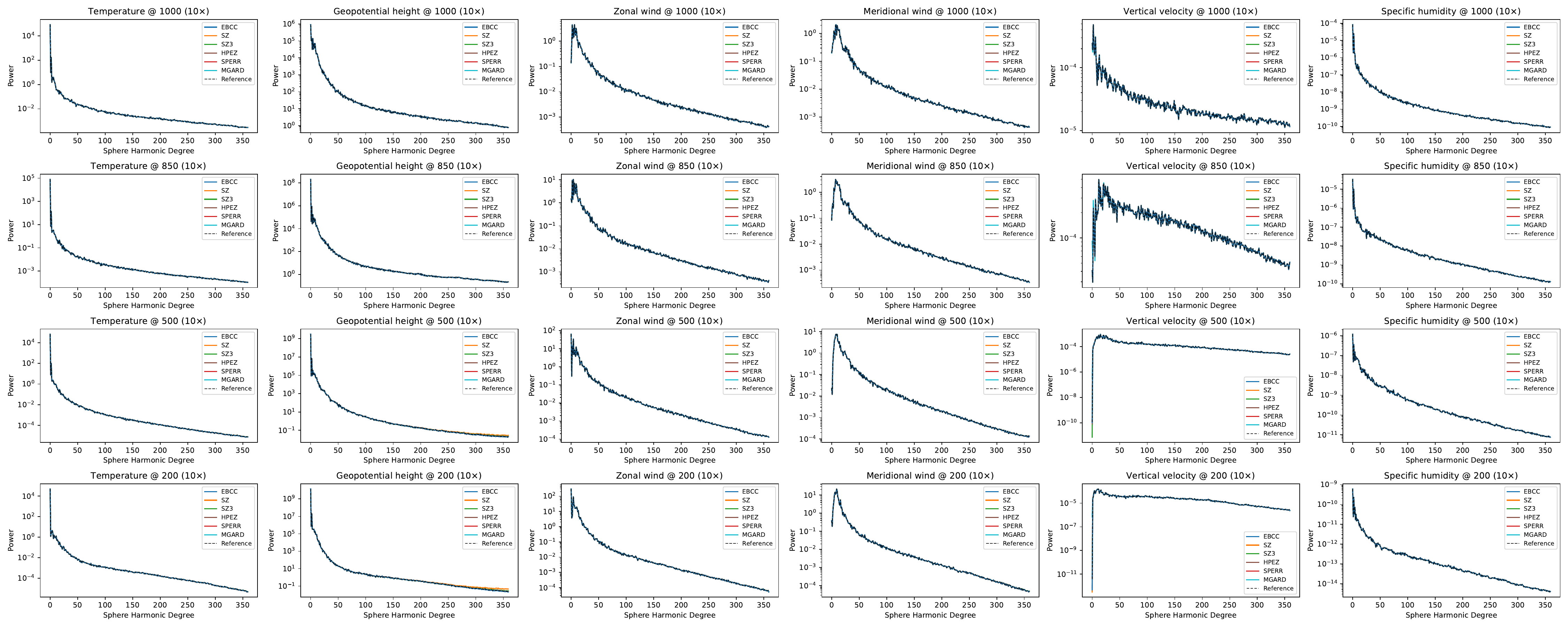}
\caption{Spherical harmonic power spectra of original and reconstructed fields for pressure-level variables at a compression ratio of 10$\times$.}\label{fig:spectrum_pl_10}
\end{figure}

\begin{figure}[htbp]
\centering
\includegraphics[width=\linewidth]{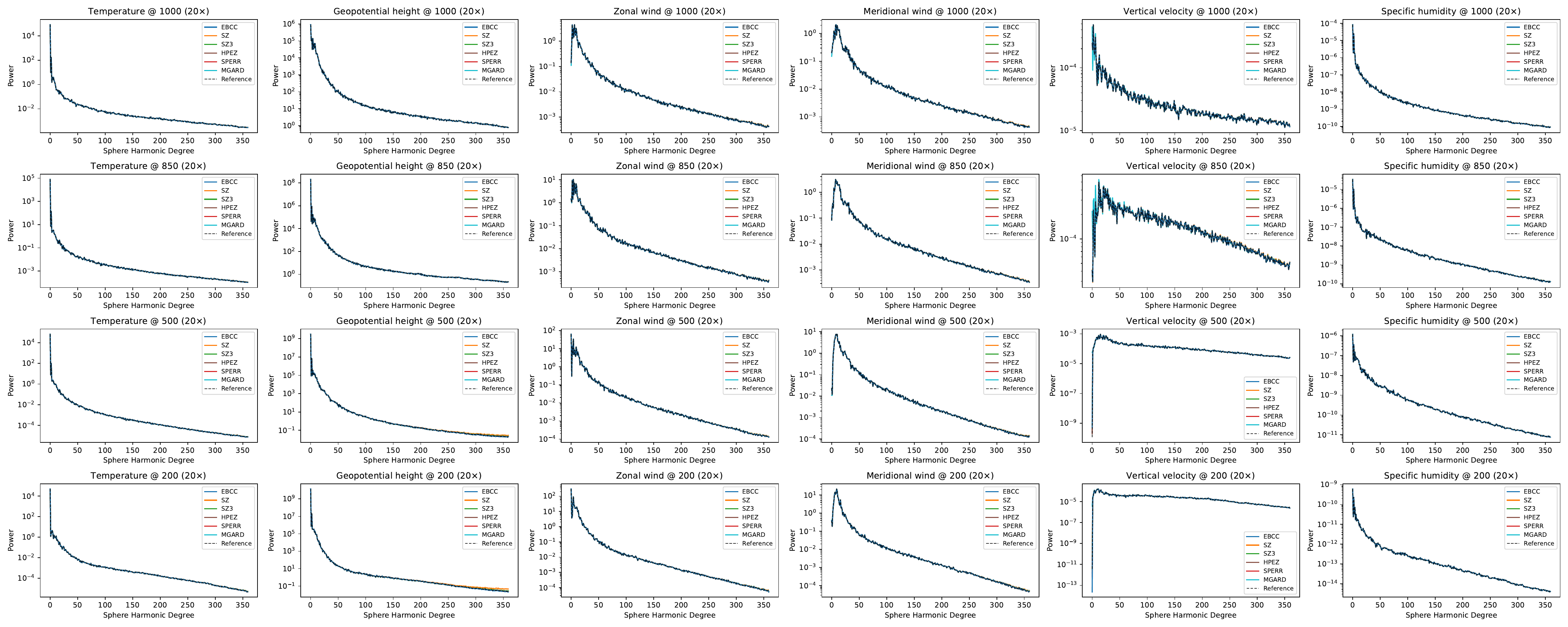}
\caption{Spherical harmonic power spectra of original and reconstructed fields for pressure-level variables at a compression ratio of 20$\times$.}\label{fig:spectrum_pl_20}
\end{figure}

\begin{figure}[htbp]
\centering
\includegraphics[width=\linewidth]{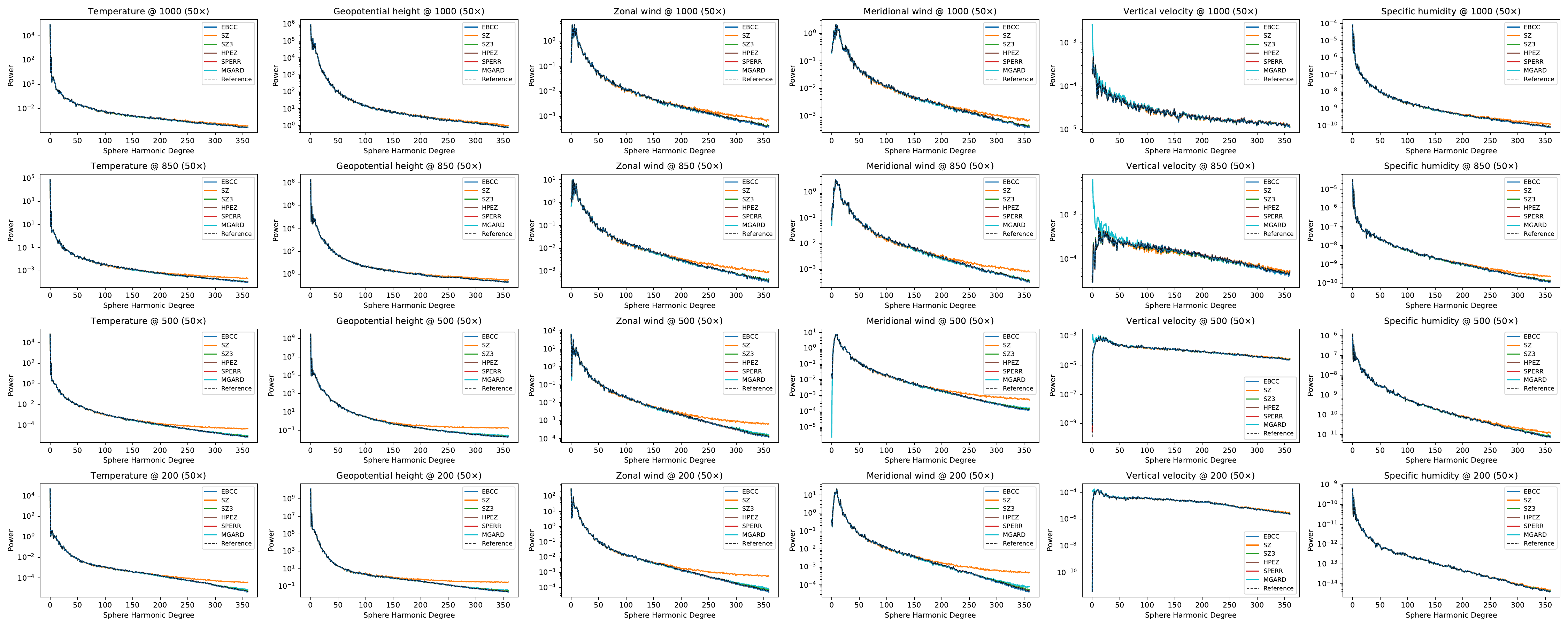}
\caption{Spherical harmonic power spectra of original and reconstructed fields for pressure-level variables at a compression ratio of 50$\times$.}\label{fig:spectrum_pl_50}
\end{figure}

\begin{figure}[htbp]
\centering
\includegraphics[width=\linewidth]{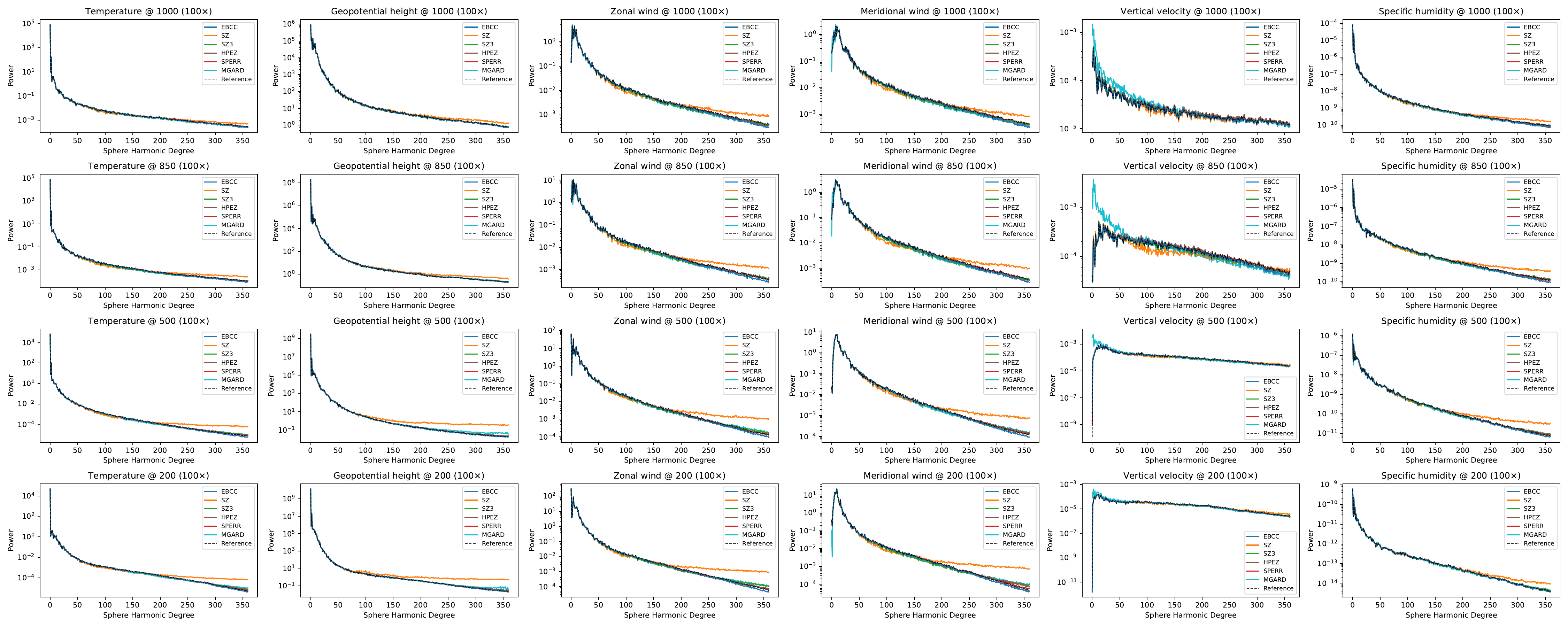}
\caption{Spherical harmonic power spectra of original and reconstructed fields for pressure-level variables at a compression ratio of 100$\times$.}\label{fig:spectrum_pl_100}
\end{figure}

\begin{figure}[htbp]
\centering
\includegraphics[width=\linewidth]{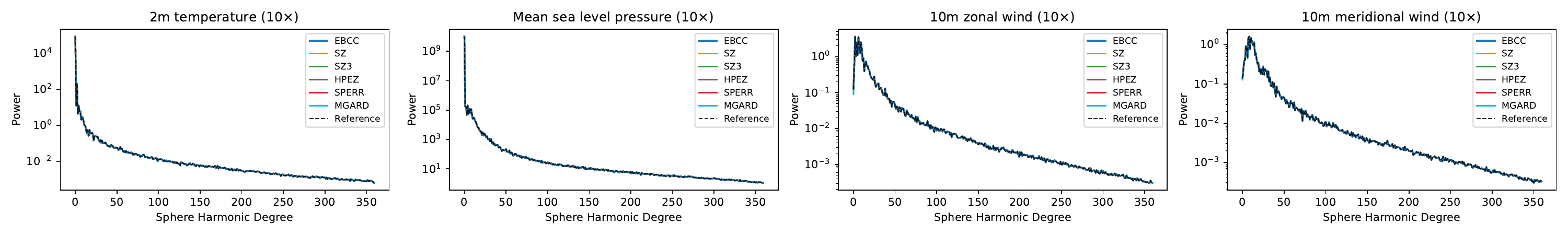}
\includegraphics[width=\linewidth]{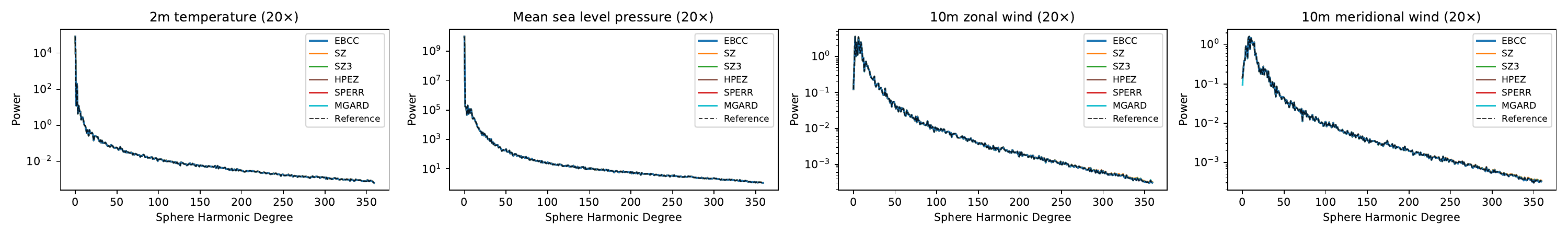}
\includegraphics[width=\linewidth]{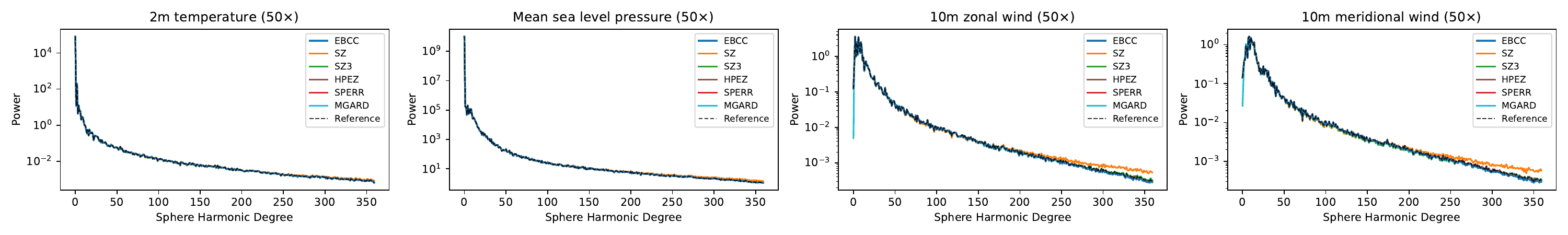}
\includegraphics[width=\linewidth]{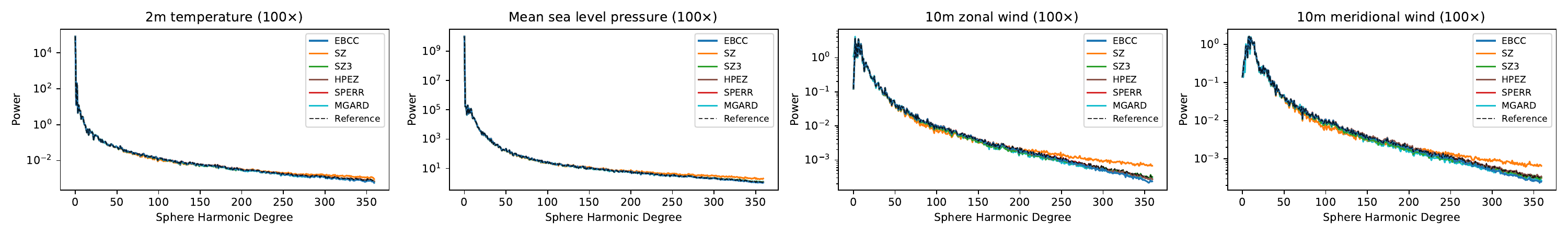}
\caption{Spherical harmonic power spectra of original and reconstructed fields for surface variables at compression ratios of 10$\times$, 20$\times$, 50$\times$, and 100$\times$ from top to bottom.}\label{fig:spectrum_sfc}
\end{figure}

\clearpage
\subsection{Case Study of Hurricane Iota}\label{app:casestudy}

\begin{figure}[htbp]
    \centering
    \includegraphics[width=\linewidth]{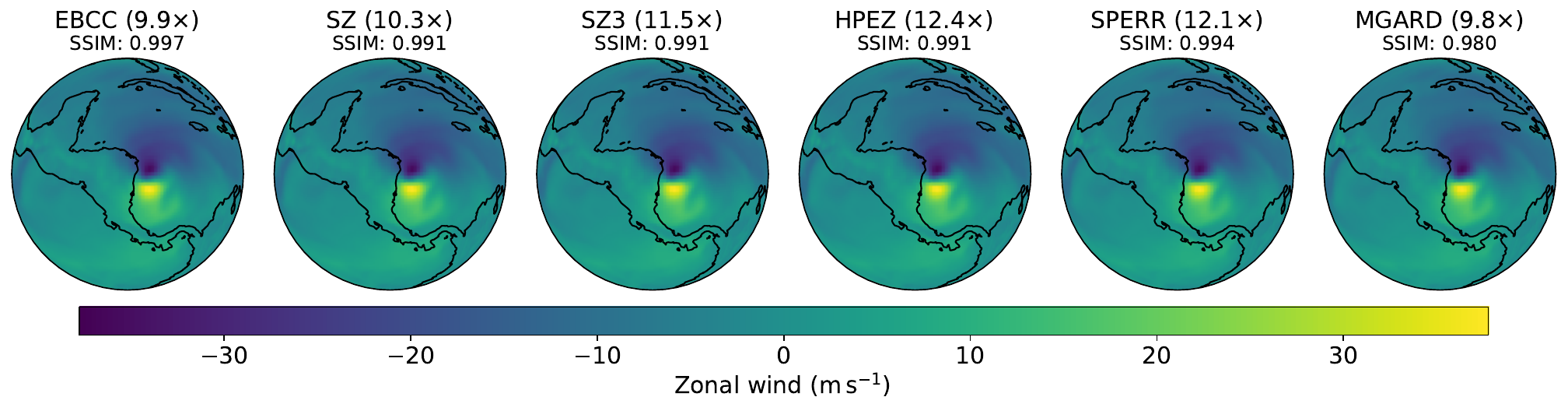}
    \includegraphics[width=\linewidth]{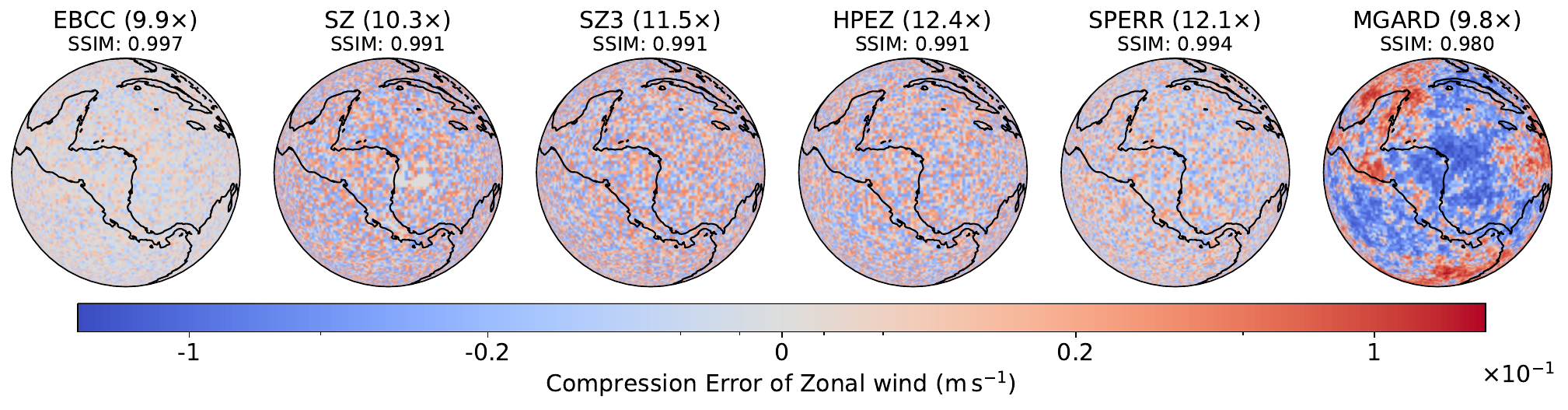}
    \includegraphics[width=\linewidth]{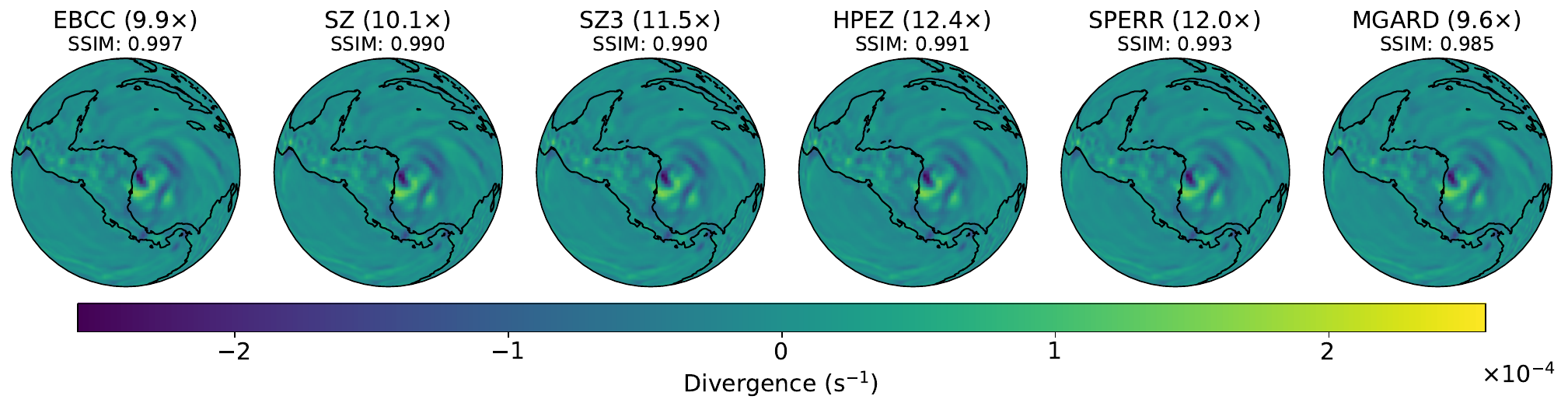}
    \includegraphics[width=\linewidth]{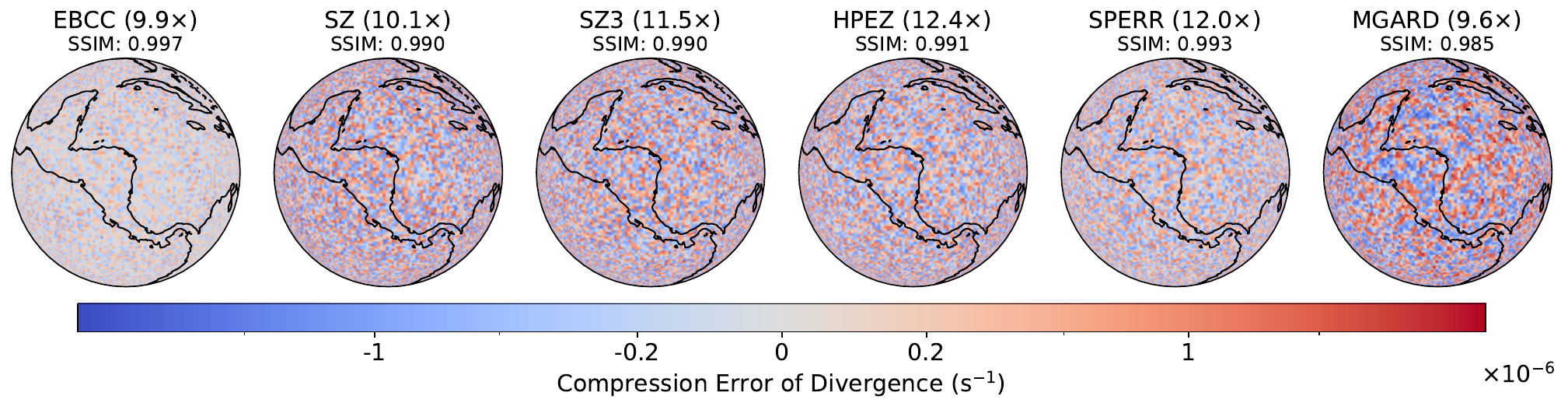}
    \caption{\textbf{Rows 1 and 3}: Plots of compressed 850hPa zonal wind speed (u) and derived horizontal wind divergence, respectively, for Hurricane Iota with compression ratios around 10$\times$. \textbf{Rows 2 and 4}: Compression-error plots corresponding to the row above. Achieved compression ratios are marked beside the method names, with Data SSIM scores calculated on the plotted region.}
    \label{fig:casestudy_iota_10}\label{fig:casestudy_iota_u850}
\end{figure}

\clearpage

\begin{figure}[htbp]
    \centering
    \includegraphics[width=\linewidth]{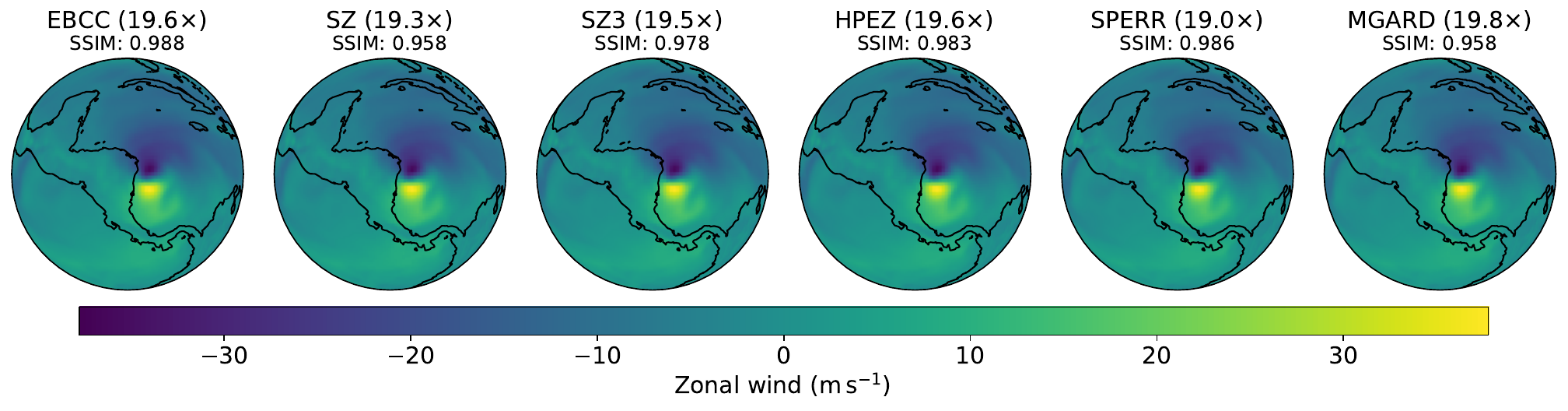}
    \includegraphics[width=\linewidth]{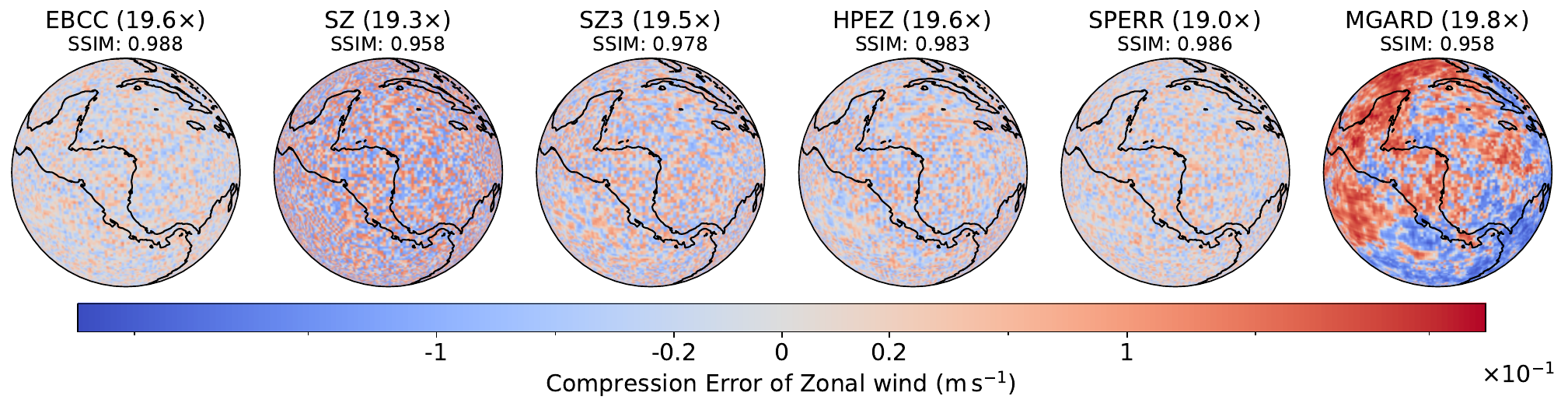}
    \includegraphics[width=\linewidth]{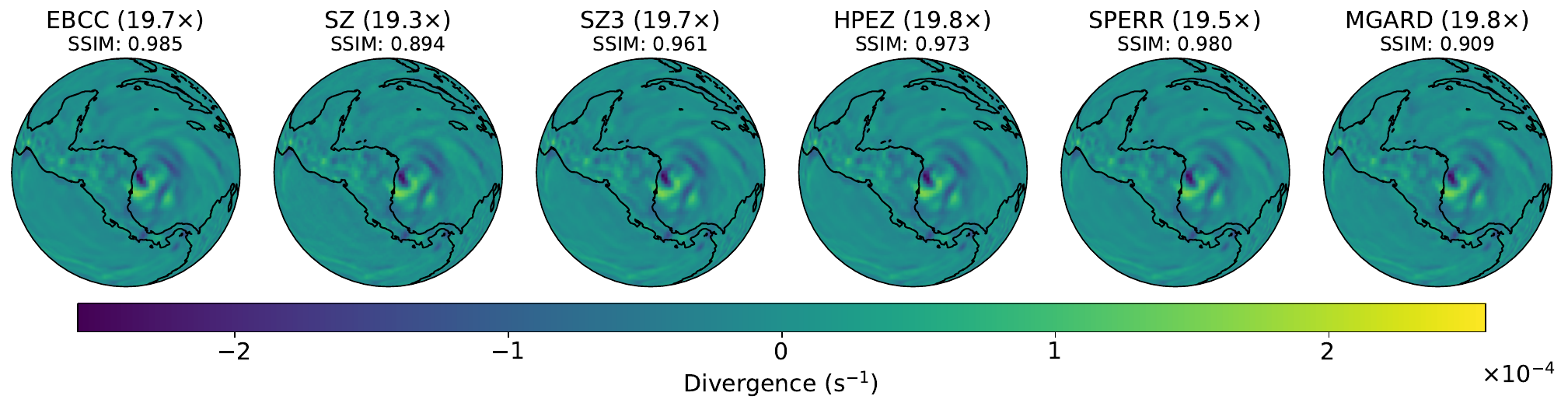}
    \includegraphics[width=\linewidth]{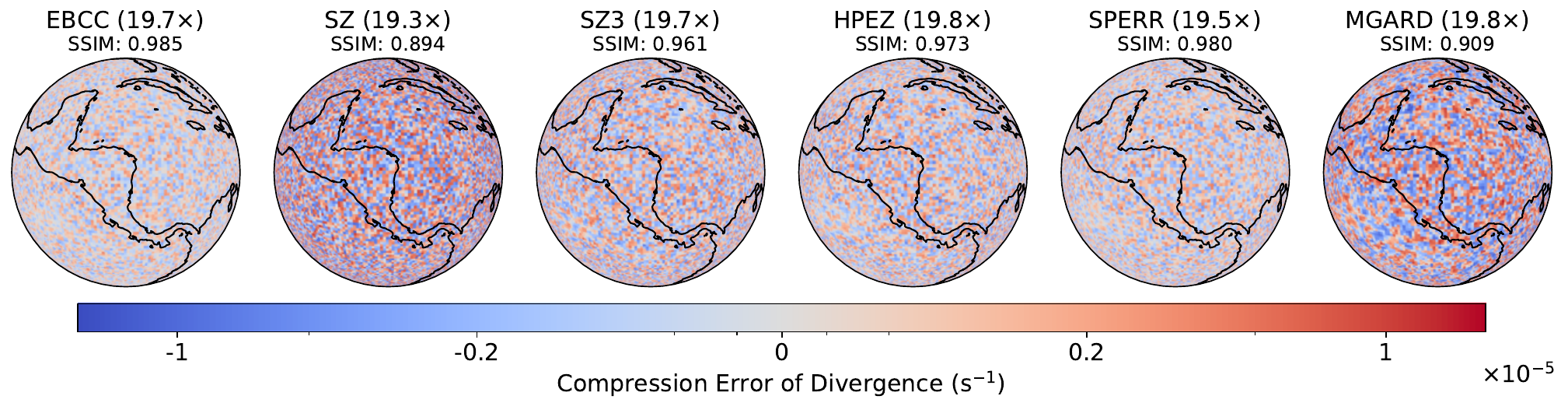}
    \caption{\textbf{Rows 1 and 3}: Plots of compressed 850hPa zonal wind speed (u) and derived horizontal wind divergence, respectively, for Hurricane Iota with compression ratios around 20$\times$. \textbf{Rows 2 and 4}: Compression-error plots corresponding to the row above. Achieved compression ratios are marked beside the method names, with Data SSIM scores calculated on the plotted region.}
    \label{fig:casestudy_iota_20}
\end{figure}

\clearpage

\begin{figure}[htbp]
    \centering
    \includegraphics[width=\linewidth]{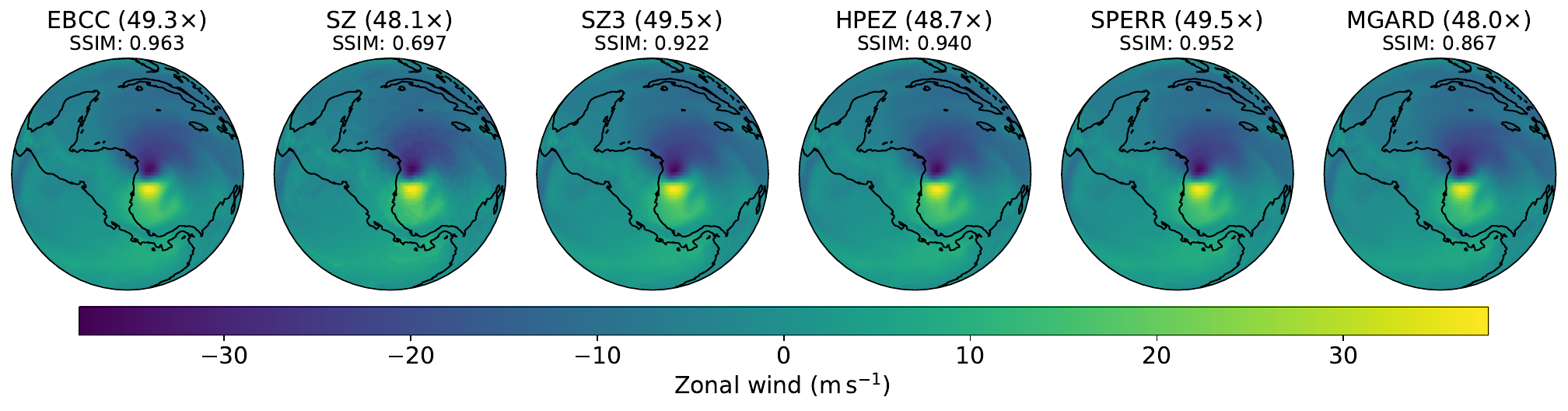}
    \includegraphics[width=\linewidth]{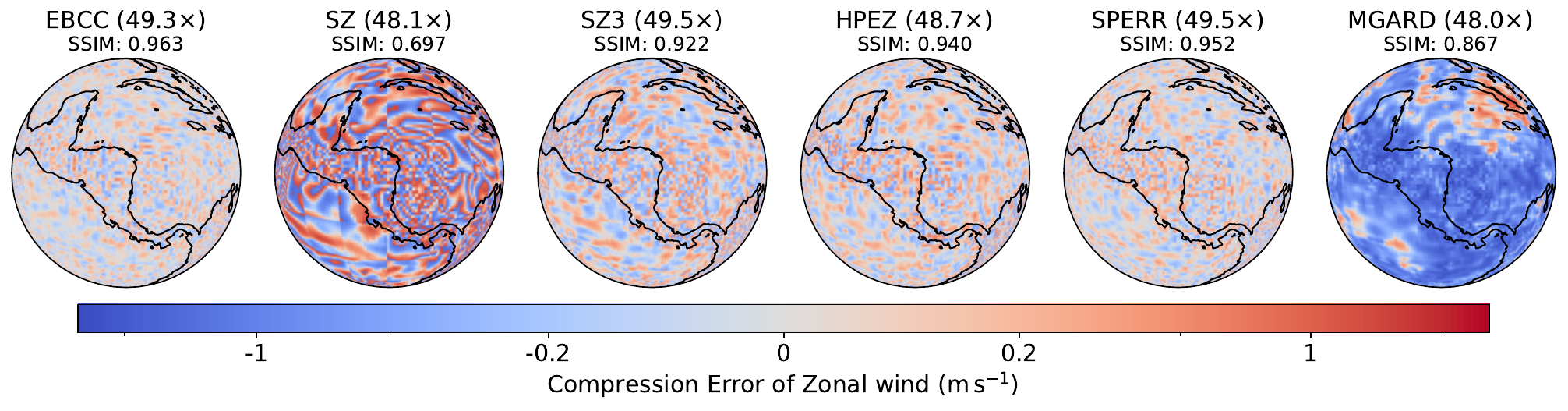}
    \includegraphics[width=\linewidth]{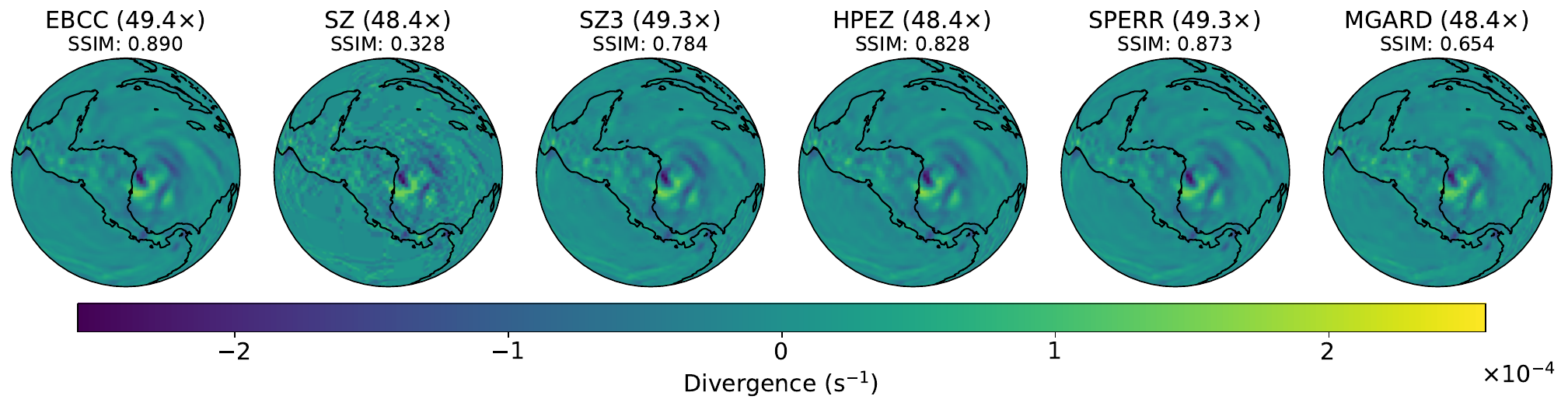}
    \includegraphics[width=\linewidth]{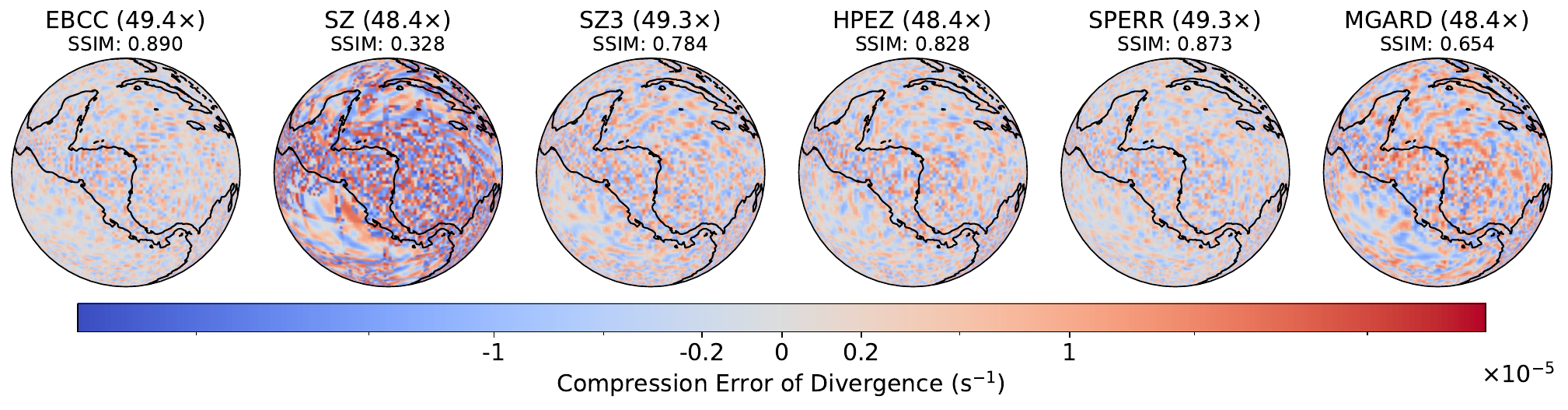}
    \caption{\textbf{Rows 1 and 3}: Plots of compressed 850hPa zonal wind speed (u) and derived horizontal wind divergence, respectively, for Hurricane Iota with compression ratios around 50$\times$. \textbf{Rows 2 and 4}: Compression-error plots corresponding to the row above. Achieved compression ratios are marked beside the method names, with Data SSIM scores calculated on the plotted region.}
    \label{fig:casestudy_iota_50}\label{fig:casestudy_iota_d850}
\end{figure}

\clearpage
\section{Software environment}\label{app:software_environment}
\autoref{tab:compressor_versions} lists versions of the compressors and supporting software in the public benchmark environment. We use the SZ, SZ3, and SPERR libraries bundled with hdf5plugin 6.0.0. We apply a bug-fix patch distributed with the benchmark source to HPEZ.

\begin{table}[htbp]
\centering
\caption{Compressor and supporting software versions. Versions are those declared by the source.}\label{tab:compressor_versions}
\begin{tabular}{ll}
\tophline
\textbf{Software} & \textbf{Version} \\ \middlehline
EBCC & 0.2.0 \\
SZ & 2.1.6.2 \\
SZ3 & 3.1.7 \\
HPEZ & 2.0.0.0 (patched) \\
SPERR & 0.8.4 \\
MGARD & 1.7.0 \\
\middlehline
OpenJPEG & 2.5.3 (modified) \\
Zstandard & 1.5.6 \\
hdf5plugin & 6.0.0 \\
LAGRANTO & ECMWF 2.7 (SVN revision 47) \\
\bottomhline
\end{tabular}
\end{table}

\noappendix       
\authorcontribution{L.H., L.F., F.S., A.C., T.H. contributed to the implementation and experiment design. J.Z., M.A.S., S.S. contributed to the experiment design.}

\competinginterests{The authors declare there are no conflicts of interest for this manuscript.} 

\begin{acknowledgements}
This work received funding from the WeatherGenerator project (grant agreement No 101187947). It was supported under project ID a01 and a122 as part of the Swiss AI Initiative, through a grant from the ETH Domain and computational resources provided by the Swiss National Supercomputing Centre (CSCS) under the Alps infrastructure. We also thank CSCS for the provisioning of compute resources for this project. The authors declare the use of generative AI in the research and writing process. According to the GAIDeT taxonomy \citep{suchikova2026gaidet}, the following tasks were delegated to GAI tools under full human supervision: code generation, process automation, data collection, data curation and organization, data analysis, visualization, and proofreading and editing. The GAI tools used were: Claude Opus 4.5 -- 5.5, ChatGPT 5.3 -- 6.1 .
\end{acknowledgements}

\bibliographystyle{copernicus}
\bibliography{ref}

\end{document}